\documentclass[aps, twocolumn, twoside, superscriptaddress, nofootinbib, floatfix]{revtex4-1}

\usepackage{amsmath,amsthm}
\usepackage{amssymb,enumerate}
\usepackage{bm}
\usepackage[dvipsnames]{xcolor}
\usepackage{float}
\usepackage[T1]{fontenc}
\usepackage[pdftex]{graphicx}
\usepackage[pdftex,colorlinks=true,linkcolor=blue,citecolor=blue,urlcolor=blue]{hyperref}
\usepackage{comment}
\usepackage[caption=false]{subfig}
\usepackage{multirow}
\usepackage{tabularx, booktabs}

\newcommand{\bra}[1]{\langle #1 |}
\newcommand{\ket}[1]{| #1 \rangle}
\newcommand{\mean}[1]{\langle #1 \rangle}
\newcommand{\ceil}[1]{\left\lceil #1 \right\rceil}
\newcommand{\floor}[1]{\left\lfloor #1 \right\rfloor}

\newcolumntype{Y}{>{\centering\arraybackslash}X}

\begin{document}

\title{Solving the Hubbard model using density matrix embedding theory and the variational quantum eigensolver}

\author{Lana Mineh}
\affiliation{Phasecraft Ltd.}
\affiliation{School of Mathematics, University of Bristol}
\affiliation{Quantum Engineering Centre for Doctoral Training, University of Bristol}
\author{Ashley Montanaro}
\affiliation{Phasecraft Ltd.}
\affiliation{School of Mathematics, University of Bristol}

\date{\today}

\begin{abstract}

Calculating the ground state properties of a Hamiltonian can be mapped to the problem of finding the ground state of a smaller Hamiltonian through the use of embedding methods. These embedding techniques have the ability to drastically reduce the problem size, and hence the number of qubits required when running on a quantum computer. However, the embedding process can produce a relatively complicated Hamiltonian, leading to a more complex quantum algorithm. In this paper we carry out a detailed study into how density matrix embedding theory (DMET) could be implemented on a quantum computer to solve the Hubbard model. We consider the variational quantum eigensolver (VQE) as the solver for the embedded Hamiltonian within the DMET algorithm. We derive the exact form of the embedded Hamiltonian and use it to construct efficient ansatz circuits and measurement schemes. We conduct detailed numerical simulations up to 16 qubits, the largest to date, for a range of Hubbard model parameters and find that the combination of DMET and VQE is effective for reproducing ground state properties of the model.

\end{abstract}

\maketitle


\section{Introduction}

Simulating quantum-mechanical systems relating to quantum chemistry or solid-state physics is one of the most important problems that quantum computers are anticipated to tackle~\cite{Cao2019, McArdle2020}. Quantum computers could make it possible to solve problems that will take an exponential amount of time and memory on classical computers.

Current quantum hardware is considered to be in the noisy intermediate-scale quantum (NISQ) regime~\cite{Preskill2018}. On NISQ devices the number of qubits is too low to allow for extensive error correction and the amount of noise restricts the size of quantum circuits that can be run. This has lead to an interest in hybrid quantum-classical algorithms which, analogously to machine learning techniques, employ classical optimisation routines to find quantum circuits that best solve the problem at hand. Of these, one of the most widely used -- and the one we will be considering in this paper -- is the variational quantum eigensolver (VQE), which is an algorithm that finds the ground state of Hamiltonians~\cite{McClean2016, Cerezo2020, Bharti2021}. 

Due to the limited number of qubits on NISQ devices, embedding algorithms which reduce the size of the problem Hamiltonian could be very useful. Algorithms such as density functional theory (DFT) and dynamical mean-field theory (DMFT), which have been used for decades in the classical simulation of solid-state systems, are gaining popularity in the quantum computing community~\cite{Ma2020, Rungger2019, Sheng2021, Bauer2016}. 

In this paper we study how density matrix embedding theory (DMET)~\cite{Knizia2012, Knizia2013} can be implemented on a quantum computer with VQE. In the DMET algorithm a fragment of the original system is retained, with the rest of it being mapped to a bath that is the same size as the fragment. DMET is well suited to be used with VQE since it does not require the computation of any complicated time- or frequency-dependent quantities such as Green's functions.

There have been a number of works over the past few years that have combined DMET with VQE. Rubin~\cite{Rubin2016} investigated solving the 1D Hubbard model using a fragment containing one site with unitary coupled cluster as the VQE ansatz. Yamazaki et al.~\cite{Yamazaki2018} conducted an analysis of DMET along with other embedded techniques for alkanes using classical quantum chemistry simulations to estimate qubit counts and sampling errors. More recently there have been experiments done on quantum hardware. Kawashima et al.~\cite{Kawashima2021} conducted an experiment on a trapped-ion quantum computer using an embedded Hamiltonian with two qubits to estimate the energy of a ring of hydrogen atoms. Tilly et al.~\cite{Tilly2021} solved the Hubbard model on a Bethe lattice using energy-weighted DMET with four qubits on IBM superconducting hardware.

In this paper we aim to go beyond these small-scale experiments with a more systematic study into the use of DMET with VQE, using the Hubbard model as a test case. The Hubbard model~\cite{Hubbard1963} is one of the simplest models of interacting electrons in a grid. The 2D case has remained unsolved and it is thought to be relevant to applications such as high-temperature superconductivity~\cite{hubbard}. Previous numerical simulations, and experiments on quantum hardware, of DMET with VQE have been limited to a fragment of one site. Here we consider fragment sizes of up to four sites (16 qubits) for solving the 1D and 2D Hubbard models, enabling us to draw conclusions about the likely scaling of DMET for larger problem sizes. Our numerical simulations include the use of measurements that take into account statistical noise.

We develop efficient algorithms based on the use of fermionic swap networks~\cite{Kivlichan2018} to implement the VQE ansatz circuit, and efficient procedures for reducing the number of measurement rounds needed. These enable us to give the first full quantum circuit complexity analysis of DMET with VQE. These results are given in Table~\ref{tab:theory_layers} for combinations of the Hubbard model dimension and rectangular shaped fragments.

\begin{table}[tb]
\centering
\begin{tabularx}{\linewidth}{ccYY}
\toprule
Hubbard             & Fragment            & Ansatz depth                          & Measurements      \\ \midrule
1D                  & 1D                  & $N_\text{frag}+3$                     & $N_\text{frag}+2$ \\ \addlinespace[0.5em]
2D                  & 1D                  & $2N_\text{frag}$                      & $2N_\text{frag}$  \\ \addlinespace[0.5em]
\multirow{2}{*}{2D} & \multirow{2}{*}{2D} & $N_\text{frag} + N_E + N_x - 2$       & \multirow{2}{*}{$N_\text{frag}+N_E$}                  \\
                    &                     & $+ (N_y - 4)\ceil{\frac{N_x - 4}{2}}$ &                   \\ \bottomrule
\end{tabularx}
\caption{Number of layers of two-qubit gates required to implement one layer of the ansatz, and circuit preparations needed to measure all of the terms in the embedded Hamiltonian. $N_\text{frag}$ is the DMET fragment size and corresponds to an embedded system with $4N_\text{frag}$ qubits. For a 2D fragment we take $N_\text{frag} = N_x \times N_y$ where we assume $N_x \leq N_y$. $N_E = 2(N_x + N_y - 2)$ is the number of sites on the edge of the 2D fragment.}
\label{tab:theory_layers}
\end{table}

We find that DMET with VQE is an efficient and accurate method for finding ground state properties of the Hubbard model. In our experiments using VQE as an approximate solver, we were able to reproduce previous results based on exact diagonalisation (see Figures~\ref{fig:energy_curves} and~\ref{fig:double_occ_curves}). However, the circuits produced using DMET are more complex than approaches based on direct truncation of the lattice~\cite{Cade2020, Cai2020}. 

For example, if we consider a quantum computer with 64 qubits then we could do a DMET calculation with 16 sites in the fragment. Taking the shape of the fragment to be $4 \times 4$, one layer of the ansatz would require a two-qubit gate depth of 30, and 32 preparations of the quantum circuit would be needed to measure all of the expectation terms. However, we could solve a $4 \times 8$ Hubbard model with open boundary conditions using a two-qubit gate depth of 9 per ansatz layer and 5 circuit preparations~\cite{Cade2020}. 

The outline of this paper is as follows. In Section~\ref{sec:dmet} we discuss the idea behind DMET, formally define the problem of the Hubbard model that we solve and lay out the steps of the variant of DMET we will be using -- the single-shot embedding algorithm~\cite{Wouters2016}. We also explicitly state what the form of the embedded Hamiltonian is (and include a derivation in Appendix~\ref{app:form_emb_hamiltonian}) which is important for the implementation of the VQE algorithm. 

In Section~\ref{sec:vqe} we briefly introduce the VQE algorithm and the Hamiltonian variational (HV) ansatz~\cite{Wecker2015}. We then present schemes involving swap networks for efficiently implementing the ansatz circuits on a quantum computer and discuss how expectation values can be measured. 

Finally in Section~\ref{sec:results} we present the results from the numerical simulations. We run simulations for a range of parameters of the Hubbard model and find that the combined DMET and VQE algorithm is effective for all of the fragment sizes tested. We reproduce graphs from the original DMET paper of Knizia and Chan~\cite{Knizia2012} and compare with exact Bethe ansatz results~\cite{Lieb1968, Shiba1972} to demonstrate that observables relevant to the Hubbard model can be calculated to a high accuracy when using VQE as the solver. Simulations involving measurements up to a fragment size of two sites (8 qubits) are also run and we discuss some of the additional complexities that can occur when running on quantum hardware.

\section{Density Matrix Embedding}
\label{sec:dmet}

The idea behind embedding methods is that the properties of a Hamiltonian $H$ can be reproduced using a smaller embedded Hamiltonian. DMET is one method for obtaining a suitable embedded Hamiltonian~\cite{Knizia2012}. 

In general, states of a quantum system can be written in terms of the basis states of two of its sub-systems. For our purposes, let us call the first sub-system $F$ the fragment and the second sub-system $E$ the environment. For example, for a system that consists of electrons in a grid, the fragment could be a subset of sites of the grid. Any state $\ket{\Psi}$ of the system can be written as
\begin{equation}
    \ket{\Psi} = \sum_{i=1}^{N_F} \sum_{j=1}^{N_E} \Psi_{ij} \ket{F_i}\ket{E_j},
\end{equation}
where $\ket{F_i}, \ket{E_j}$ are basis states of $F$ and $E$, and $N_{F/E}$ are the sizes of their respective Hilbert spaces. Using the singular value decomposition for $\Psi_{ij}$ it can be rewritten as
\begin{align}
    \ket{\Psi} &= \sum_{i=1}^{N_F} \sum_{j=1}^{N_E} \sum_{\alpha=1}^{\text{min}(N_F, N_E)} U_{i\alpha} \lambda_\alpha V^\dagger_{\alpha j} \ket{F_i}\ket{E_j} \nonumber\\
    &= \sum_{\alpha=1}^{N_F} \lambda_\alpha \ket{F'_\alpha}\ket{B_\alpha},
\end{align}
where without loss of generality we have taken $N_E > N_F$. The $\ket{F_i}$ states have been rotated to a new basis $\ket{F'_\alpha} = \sum_i U_{i\alpha} \ket{F_i}$ of the fragment. The $\ket{B_\alpha} = \sum_j V^\dagger_{\alpha j} \ket{E_j}$ are a reduced set of states, called the bath, which represent the portion of the environment needed to model interactions with the fragment. This is the Schmidt decomposition of $\ket{\Psi}$~\cite{Wouters2016}.

If $\ket{\Psi}$ were the ground state of a Hamiltonian $H$ in the full system, then by construction it is also the ground state of a smaller embedded Hamiltonian $H_{\text{emb}}$ given by
\begin{equation} \label{eq:H_emb_theory}
    H_{\text{emb}} = \mathcal{P}^\dagger H  \mathcal{P},
\end{equation}
with the projector $ \mathcal{P}$ being
\begin{equation}
     \mathcal{P} = \sum_{\alpha\beta} \ket{F'_\alpha B_\beta} \bra{F'_\alpha B_\beta}.
\end{equation}

In practice $\ket{\Psi}$ is not known so the exact embedding procedure cannot take place. Instead we look to approximate $\mathcal{P}$ by taking the Schmidt decomposition of another state $\ket{\Phi}$ which is determined self-consistently. Typically, $\ket{\Phi}$ is taken to be the ground state of a mean-field quadratic Hamiltonian $H_{MF}$, where $H_{MF}$ is an approximation to $H$, as this can be calculated efficiently. Furthermore, depending on the variant of DMET, an alternative to to equation~(\ref{eq:H_emb_theory}) may be used to determine the embedded Hamiltonian from $\mathcal{P}$.

At the end of the self-consistency procedure, observables of $H_\text{emb}$ are used to approximate observables of the full Hamiltonian $H$. This is described in Section~\ref{sec:observables}.

\subsection{Single-shot embedding for the Hubbard model}
\label{sec:single_shot_embedding}

There are many variants of the DMET procedure which choose different mean-field Hamiltonians $H_{MF}$, different ways of projecting onto the problem Hamiltonian $H$ and different termination criteria for self-consistency. Here we have chosen to focus on the simplest form of DMET, single-shot embedding~\cite{Bulik2014_2, Wouters2016}. This will highlight the key issues that would be associated with implementing any form of DMET on a quantum computer. Single-shot embedding has been shown to be effective in practice~\cite{Bulik2014_2, Yamazaki2018} (see Appendix~\ref{app:single_shot_embedding}) and has been successfully used with the VQE algorithm~\cite{Rubin2016, Kawashima2021} for one fragment site. 

Here we will briefly lay out the steps in the single-shot embedding algorithm in the context of the Hubbard model. A more detailed explanation is given in Appendix~\ref{app:single_shot_embedding}. The Hubbard Hamiltonian is defined as 
\begin{equation} 
    H_\text{hub} = -t \sum_{\mean{i, j}, \sigma} a^\dagger_{i\sigma} a_{j\sigma} + U \sum_i n_{i\uparrow} n_{i\downarrow}
    = T + W,
\end{equation}
where $a^\dagger_{i\sigma}$ and $a_{i\sigma}$ are the creation and annihilation operators for a spin $\sigma \in \{\uparrow, \downarrow\}$ fermion in site $i$, and $n_{i\sigma} = a^\dagger_{i\sigma} a_{i\sigma}$. The notation $\mean{i, j}$ indicates that the sum is performed over neighbouring sites $i, j$ in the grid. $T$ describes the kinetic energy in the system; it contains the single-particle hopping terms with $t$ being the tunnelling amplitude. $W$ describes the interactions between particles in the system. It is often called the onsite term and $U$ is the Coulomb potential. 

We will be considering the problem of finding properties of the ground state of the model on an infinite 1D or 2D rectangular grid that has a fixed fraction of the sites filled with electrons, with the same proportion of up and down. In practice we will approximate the infinite grid by a large number of sites $N$ with periodic or anti-periodic boundary conditions\footnote{Periodic boundary conditions introduce $-ta^\dagger_{0\sigma}a_{N\sigma}$ terms into the Hamiltonian, anti-periodic boundary conditions introduce $ta^\dagger_{0\sigma}a_{N\sigma}$.}, occupied by $N_\text{occ}$ fermions split equally between up and down. The procedure to reduce this problem to an embedded Hamiltonian with $N_\text{frag}$ sites in the fragment using single-shot embedding is as follows~\cite{Wouters2016}:

\begin{enumerate}
    \item Calculate the ground state of the approximating mean-field Hamiltonian which in this case is taken to be the quadratic part of $H_\text{hub}$, $H_{MF} = T$. $H_{MF}$ can be solved efficiently and its ground state $\ket{\Phi}$ is a Slater determinant.
    
    \item Construct the projector $ \mathcal{P}$ from the one-particle reduced density matrix (1-RDM) of $\ket{\Phi}$. These first two steps are equivalent to taking the Schmidt decomposition of $\ket{\Phi}$ as described in the previous section.
    
    \item Use the projector to construct the embedded Hamiltonian
    \begin{equation} \label{eq:H_emb}
        H_{\text{emb}} = T_\text{emb} + W_\text{emb} - \mu \sum_{i \in \text{frag}, \sigma} n_{i\sigma}.
    \end{equation}
    We use the non-interacting bath formulation to construct the embedded Hamiltonian from $\mathcal{P}$. This involves only projecting the quadratic part of $H_\text{hub}$, $T_\text{emb} = \mathcal{P}^\dagger T \mathcal{P}$ and taking $W_\text{emb}$ to be the terms in $W$ that act on the fragment. $\mu$ is an added chemical potential term that governs the number of electrons in the fragment -- the ``single-shot'' refers to this single free parameter.
    
    \item Solve the embedded problem $H_\text{emb}$ which is a Hamiltonian on $4N_\text{frag}$ orbitals ($2N_\text{frag}$ for each spin's fragment and bath sites). The ground state $\ket{\Phi_\text{emb}}$ of $H_\text{emb}$ can be found using methods such as exact diagonalisation, DMRG, or VQE.
    
    \item Repeat from step 3, adjusting the chemical potential $\mu$ until the fraction of occupied orbitals in the fragment matches the site occupancy of $H_\text{hub}$.
\end{enumerate}

More general forms of DMET have an extra optimisation loop. A correlation potential $V$ is introduced in the mean-field Hamiltonian, giving $H_{MF} = T + V$, which is adjusted until the 1-RDMs of $\ket{\Phi}$ and $\ket{\Phi_\text{emb}}$ match~\cite{Wouters2016}.

\subsection{Calculating observables from the embedded Hamiltonian}
\label{sec:observables}

Observables relevant to the original problem Hamiltonian $H_\text{hub}$ can be calculated from the final $\ket{\Phi_\text{emb}}$ given by the DMET algorithm. The quantities of interest in this paper are the energy and double occupancy per site. These are calculated by taking expectation values of $\ket{\Phi_\text{emb}}$ on the fragment and fragment-bath. Contributions purely from the bath are ignored.

For example, the energy of the fragment is calculated as~\cite{Zheng2016, Knizia2013}
\begin{equation}
    E_\text{frag} = \mean{T_\text{emb}^\text{frag}} + \frac{1}{2}\mean{T_\text{emb}^\text{frag-bath}} + \mean{W_\text{emb}},
\end{equation}
where $T_\text{emb}^\text{frag}$ and $T_\text{emb}^\text{frag-bath}$ are the terms of $T_\text{emb}$ that act on the fragment-only, or between the fragment and bath, respectively. The energy per site is then obtained by dividing by the number of sites in the fragment. Double occupancy of the fragment is calculated as~\cite{Bulik2014}
\begin{equation}
    D_\text{frag} = \sum_{i \in \text{frag}} \mean{n_{i\uparrow}n_{i\downarrow}} = \frac{\mean{W_\text{emb}}}{U}.
\end{equation}

\subsection{Form of the embedded Hamiltonian}
\label{sec:form_emb_hamiltonian}

This section contains a summary of the structure of the embedded Hamiltonian, which will be necessary for developing efficient swap networks and measurement schemes in Section~\ref{sec:vqe}. Unlike when using a classical procedure such as exact diagonalisation, having more terms in the embedded Hamiltonian results in a more complicated circuit being run on the quantum computer and requires more measurements to estimate the expectation values.

In general, the embedded Hamiltonian from equation~(\ref{eq:H_emb}) can be written explicitly as
\begin{align}
    H_\text{emb} &= \sum_{i, j \in \text{emb}, i<j, \sigma} t_{ij} (a^\dagger_{i\sigma} a_{j\sigma} + a^\dagger_{j\sigma} a_{i\sigma}) \nonumber\\
    &+ \sum_{i \in \text{bath}, \sigma} t_{ii} n_{i\sigma} + U \sum_{i \in \text{frag}} n_{i\uparrow} n_{i\downarrow} - \mu \sum_{i \in \text{frag}, \sigma} n_{i\sigma}.
\end{align}

Determining the form of $H_\text{emb}$ now comes down to knowing which terms are present in the Hamiltonian (non-zero $t_{ij}$). Here we will state the structure of the embedded Hamiltonian when the single-shot embedding procedure is carried out for the 1D and 2D Hubbard models. These results are derived by considering the matrix of coefficients of $T$ and its projection $T_\text{emb}$. This derivation has been made possible due to the simple structure of $T$ and properties of the Hubbard model such as translational invariance. The structure may be difficult to calculate for a general quadratic Hamiltonian. Details are provided in Appendix~\ref{app:form_emb_hamiltonian}. 

There are three different types of terms to consider -- fragment-only terms, bath-only terms and fragment-bath hopping terms. The fragment-only terms retain the same structure as the Hubbard model and are nearest neighbour hopping terms. For the fragment-bath interactions, each fragment site on the edge of the fragment shares a hopping term with all of the bath sites. In the 1D Hubbard model, the edge sites are the first and last sites of the fragment. For the 2D Hubbard model with a 1D fragment all of the fragment sites are on the edge. 

For the terms acting only on the bath, the $t_{ii}$ are generally all non-zero. When using anti-periodic boundary conditions with the Hubbard model and taking the number of electrons of one spin type $(N_\text{occ}/2)$ to be even (or with periodic boundary conditions and $N_\text{occ}/2$ odd), the bath hopping terms in $H_\text{emb}$ split into two groups -- even and odd numbered sites. Within each of these two groups, every site has a hopping term with all the other sites. If these conditions are not met then all of the bath sites can interact with all of the other bath sites, increasing the number of interactions.

The embedded Hamiltonian of the 2D Hubbard model with a 1D fragment has the same structure of bath hopping terms as the 1D model. However, when using a 2D fragment, the bath sites split into four groups where within each group all possible interactions occur. Unlike the 1D case there is no clear split (e.g. even/odd), but the size of the groups are roughly equal. Conditions on when this split into four groups occurs is discussed in Appendix~\ref{app:form_emb_hamiltonian}. 

From the standpoint of quantum circuit complexity, it will always be advantageous to use a 2D shaped fragment when solving the 2D Hubbard model. This is due to the fact that both the fragment-bath and bath-only hopping terms will be fewer in number than when using a 1D fragment shape.

\section{Variational Quantum Eigensolver}
\label{sec:vqe}

The VQE is a hybrid quantum-classical algorithm used to produce the ground state of a Hamiltonian $H$. It relies on the variational principle, which states that $\bra{\psi} H \ket{\psi} \geq E_g$, where $\ket{\psi}$ is an arbitrary normalised state and $E_g$ is the ground energy of $H$. The steps of the algorithm are~\cite{McClean2016, Cerezo2020}:

\begin{enumerate}
    \item Prepare a parameterised state $\ket{\psi(\bm{\theta})} = U(\bm{\theta}) \ket{\psi_0}$ on the quantum computer. $U(\bm{\theta})$ is an ansatz circuit intended to reproduce the ground state and $\ket{\psi_0}$ is an initial starting state.
    
    \item Measure the expectation value $\bra{\psi(\bm{\theta})} H \ket{\psi(\bm{\theta})}$.
    
    \item Use a classical optimisation method to determine a new value for $\bm{\theta}$ that will minimise the expectation value.
    
    \item Repeat steps 1-3 until the optimiser converges. The final value of $\bm{\theta}$ will parameterise the ground state and give an expectation value equal to the ground energy.
\end{enumerate}

We will be using the VQE algorithm to find the ground state of $H_\text{emb}$ in equation~(\ref{eq:H_emb}). $H_\text{emb}$ is a fermionic Hamiltonian and must first be expressed as a qubit Hamiltonian. We use the Jordan-Wigner encoding~\cite{Somma2002} which introduces no overhead in qubit count, as each orbital maps to one qubit. Note that the Jordan-Wigner encoding requires us to choose an ordering for the orbitals as it maps the fermionic modes to a line (e.g.\ order the fragment sites before the bath sites and all the up orbitals before the down). 

The hopping terms between qubits are transformed as
\begin{equation}
    a^\dagger_i a_j + a^\dagger_j a_i \mapsto \frac{1}{2} (X_i X_j + Y_i Y_j) Z_{i+1} \cdots Z_{j-1},
\end{equation}
where $i < j$ without loss of generality, and $X_i, Y_i, Z_i$ are the Pauli matrices acting on qubit $i$. The number operator terms become
\begin{equation}
    \label{eq:jw_number}
    n_i = a^\dagger_i a_i \mapsto \frac{1}{2} (I - Z_i) = \ket{1}\bra{1}_i,
\end{equation}
where $I$ is the identity matrix, and the onsite terms become
\begin{equation}
    \label{eq:jw_onsite}
    n_i n_j \mapsto \frac{1}{4} (I - Z_i)(I - Z_j) = \ket{11}\bra{11}_{ij}.
\end{equation}

\subsection{Ansatz circuit implementation using swap networks}

In this paper we implement the Hamiltonian variational ansatz~\cite{Wecker2015} which has been shown to be effective for the Hubbard model~\cite{Cade2020, Wecker2015, Reiner2019, Cai2020}. 

The initial state $\ket{\psi_0}$ for this ansatz is the ground state of the non-interacting ($U=0$) part of $H_\text{emb}$. This is a quadratic Hamiltonian which means its ground state can be prepared efficiently on a quantum computer using Givens rotations~\cite{Jiang2018}. $H_\text{emb}$ can be split up as $H_\text{emb} = \sum_j H_j$ where the terms inside each $H_j$ are commuting. The ansatz consists of applying evolutions of the form $e^{i\theta H_j}$ to the starting state, where $\theta$ is a parameter to be determined in the VQE optimisation loop. The parameterised state is
\begin{equation}
    \ket{\psi(\bm{\theta})} = \prod_d \prod_j e^{i \theta_{d, j} H_j} \ket{\psi_0},
\end{equation}
where applying all the $H_j$ evolutions makes up one layer of the ansatz whose depth is indexed by $d$. The purpose of $H_j$ is to group together terms which will share the same variational parameter $\theta$.

In Section~\ref{sec:results} we will be considering the two extremes of splitting $H_\text{emb}$ into commuting groups $H_j$. In one case we pick the groups so that there are as few as possible. In the other, each $H_j$ will contain only one term from $H_\text{emb}$ which leads to the maximum number of parameters per layer. This has the effect of making the optimisation routine more difficult but can reduce the ansatz depth required to solve the problem.

Moving onto the implementation of $e^{i\theta H_j}$ in terms of quantum gates, there are three types of evolutions: hopping terms, onsite terms and number operator terms. It is important that the terms in $H_j$ commute so that the quantum circuit can be decomposed into these three types of operations. The number terms $e^{i\theta n_j}$ can be implemented as a phase shift on qubit $j$ and the onsite terms $e^{i\theta n_j n_k}$ as a controlled phase shift between qubits $j$ and $k$. The hopping gate $e^{i\theta(a^\dagger_j a_k + a^\dagger_k a_j)}$ is a $k-j+1$ qubit gate due to the $Z$ strings in the Jordan-Wigner encoding. It can be decomposed as $2(k-j) + 1$ two-qubit gates -- a series of controlled-Z (CZ) gates between qubit $i$ and $k$ for $i = j$ up to $i = k-1$, followed by a number preserving rotation gate
\begin{equation}
    \text{Hopping}(\theta) = \begin{pmatrix}
    1 & 0 & 0 & 0 \\
    0 & \cos{\theta} & i\sin{\theta} & 0 \\
    0 & i\sin{\theta} & \cos{\theta} & 0 \\
    0 & 0 & 0 & 1
    \end{pmatrix}
\end{equation}
between qubits $j$ and $k$ and the same CZ gates repeated in reverse~\cite{Reiner2019}. 

This need for a relatively large numbers of gates to implement hopping gates motivates the use of fermionic swap networks~\cite{Kivlichan2018}. FSWAP gates\footnote{The action of FSWAP on the basis states is $\ket{00} \mapsto \ket{00}, \ket{01} \mapsto \ket{10}, \ket{10} \mapsto \ket{01}$ and $\ket{11} \mapsto -\ket{11}$.} are used to swap qubits around, changing the Jordan-Wigner ordering. Now, for example, if qubits $j$ and $k$ are swapped so that they are adjacent to each other in the ordering, the hopping gate between them can be implemented more simply as a two-qubit gate. 

A particular example of a swap network described in~\cite{Kivlichan2018} is one that places every qubit adjacent to every other qubit once. For $n$ qubits, the entire network can be implemented in $n$ layers of FSWAP gates. Furthermore, hopping gates can be folded into this swap network~\cite{Cade2020}, since
\begin{equation}
    \text{FSWAP} \cdot \text{Hopping}(\theta) = (S^\dagger \otimes S^\dagger) \cdot \text{Hopping}(\theta + \pi/2),
\end{equation}
where the $S$ gate implements a $\pi/2$ phase shift. In total $n$ layers of two-qubit gates plus some one-qubit corrections are required to implement all possible hopping interactions between every pair of qubits.

Turning back to the embedded Hamiltonian $H_\text{emb}$, we can restrict our analysis of the complexity of implementing the hopping terms to one spin type since the two spins are identical, making $n = 2N_\text{frag}$ in our case. $2N_\text{frag}$ layers of two-qubit gates is an upper bound since that accounts for all possible hopping interactions. If we look to the structure of the embedded Hamiltonian it is possible to reduce the number of layers required. 

In the rest of this section we present a scheme for the 1D model where all the hopping interactions are done in $N_\text{frag}+2$ layers, giving almost a factor of 2 improvement compared with the upper bound. The swap networks for the 2D model with 1D and 2D shaped fragments are more complex. The number of circuit layers required for these is stated in Table~\ref{tab:theory_layers} and full details of the swap networks are given in Appendix~\ref{app:2d_swap_network}.

Recall from Section~\ref{sec:form_emb_hamiltonian} that when $H_\text{hub}$ is one-dimensional, $H_\text{emb}$ has nearest-neighbour hopping terms on the fragment, the first and last fragment sites interact with all of the bath sites, and the bath sites split into odd/even groups where all the sites within a group interact with each other. Note that the lower bound on the layers of two-qubit gates required to implement all of these hopping gates is $N_\text{frag} + 1$, since each fragment site on the edge needs to interact with its one neighbouring fragment site and all $N_\text{frag}$ bath sites. 

Let $F_i$ denote fragment site $i$ and $B_i$ bath site $i$. We take the Jordan-Wigner ordering for one spin type to be one where the fragment edge sites $F_0$ and $F_{N_\text{frag}-1}$ start close to the bath, and the even/odd bath sites are placed next to each other. Let the ordering be
\begin{align} \label{eq:1d_jw_ordering}
    &F_{N_\text{frag}-3}\, F_{N_\text{frag}-4}\, \cdots F_2\, F_1\, F_0\, F_{N_\text{frag}-2}\, F_{N_\text{frag}-1} \nonumber \\
    &\quad B_1\, B_3\, \cdots B_{N_o}\,\, B_0\, B_2\, \cdots B_{N_e},
\end{align}
where if $N_\text{frag}$ is even, $N_o = N_\text{frag}-1$ and $N_e = N_\text{frag}-2$ and vice-versa if $N_\text{frag}$ is odd. 

At the first layer of the swap network the hopping term between $F_{N_\text{frag}-1}$ and $F_{N_\text{frag}-2}$ is carried out. For the following $N_\text{frag}$ layers, combined FSWAP and hopping gates are done between $F_{N_\text{frag}-1}$ and the bath sites to its right. This implements all of the hopping terms for the fragment edge site $F_{N_\text{frag}-1}$. Simultaneously, a similar procedure can be carried out for the other edge site $F_0$. At the first layer the hopping term $F_0 - F_1$ is implemented, at the second layer an FSWAP is done between $F_0$ and $F_{N_\text{frag}-2}$ to place $F_0$ next to the bath sites, and for the following $N_\text{frag}$ layers $F_0$ interacts with all the bath sites through combined FSWAP and hopping gates. Therefore, $N_\text{frag} + 2$ layers of two-qubit gates are required to implement all of the hopping gates associated to the fragment edge sites.

The remaining fragment- and bath-only hopping terms can be fitted within these $N_\text{frag} + 2$ layers. All of the fragment hopping terms can be implemented in 2 layers~\cite{Cade2020} except for $F_{N_\text{frag}-2} - F_{N_\text{frag}-3}$ which need to be placed adjacent to each other first. After the second layer of the swap network, there are $N_\text{frag}-4$ qubits between them, requiring $\ceil{(N_\text{frag} - 4)/2}$ layers of FSWAPs to bring the two sites next to each other in the Jordan-Wigner ordering. The hopping terms between the even bath sites can be implemented in $\ceil{N_\text{frag}/2}$ layers using the standard full swap network from~\cite{Kivlichan2018}. $F_{N_\text{frag}-1}$ takes $\floor{N_\text{frag}/2} + 1$ layers to interact with its neighbouring fragment site and the odd bath sites, so the entire even bath swap network can fit in these layers. Similarly, the odd bath swap network requires $\floor{N_\text{frag}/2}$ layers of two-qubit gates and these can all fit in the layers after they have interacted with $F_0$. 

\begin{figure}[tb]
    \centering
    \includegraphics[width=\linewidth]{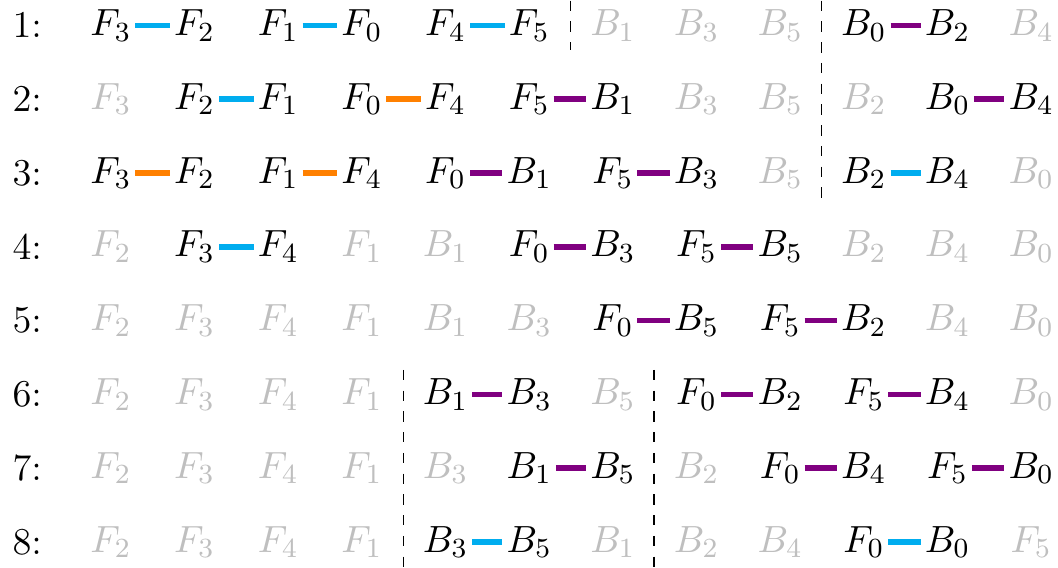}
    \caption{Demonstration of the swap network for the 1D model using fragment size 6. The blue lines are hopping gates, the orange are fermionic swap gates and the purple are combined FSWAP and hopping gates between two neighbouring qubits. The dashed lines are added to aid the eye and unused qubits are greyed out.}
    \label{fig:1Dswap_network}
\end{figure}

Figure~\ref{fig:1Dswap_network} demonstrates this full procedure with fragment size 6. Note that this swap network leaves the qubits in a less structured order. To get the same two-qubit gate depth of $N_\text{frag} + 2$, we simply reverse the swap network at the next ansatz depth. 

Finally, incorporating the onsite and number gates brings the total two-qubit gate depth for one complete layer of the ansatz to $N_\text{frag} + 3$. All the onsite gates need an additional layer to complete since they are all two-qubit gates on disjoint pairs of qubits. The number operator terms are one-qubit gates which can act on idle qubits in the swap network.

\subsection{Measuring expectation values}
\label{sec:measurement}

At the end of each run of the ansatz circuit we must measure the energy of the ansatz state $\ket{\psi(\bm{\theta})}$ with respect to $H_\text{emb}$, ideally in as few preparations of the circuit as possible. All of the onsite and number operator terms can be measured simultaneously by doing a computational basis measurement on every qubit. From the Jordan-Wigner forms given in equations~(\ref{eq:jw_number}) and~(\ref{eq:jw_onsite}), the expectation $\mean{n_i}$ is the probability of getting 1 when measuring qubit $i$ and $\mean{n_i n_j}$ is the probability of measuring a 1 on both qubit $i$ and $j$.  

Calculating the expectation values of all the hopping terms $\mean{a^\dagger_i a_j + a^\dagger_j a_i}$ requires multiple circuit preparations. To measure the hopping terms we use the method from~\cite{Cade2020}, where an operator $M$ that diagonalises $\frac{1}{2}(XX+YY)$ is applied to qubits $i$ and $j$. Such an $M$ is given by
\begin{equation}
    M = \begin{pmatrix}
        1 & 0 & 0 & 0 \\
        0 & \frac{1}{\sqrt{2}} & \frac{1}{\sqrt{2}} & 0 \\
        0 & \frac{1}{\sqrt{2}} & -\frac{1}{\sqrt{2}} & 0 \\
        0 & 0 & 0 & 1 \\
        \end{pmatrix}
\end{equation}
and can be implemented by a CNOT gate followed by a controlled-Hadamard gate and another CNOT gate. Computational basis measurements are then done on all the qubits from $i$ to $j$ and their statistics processed~\cite{Cade2020}. Alternative techniques can be used such as transforming into the Bell basis~\cite{Hamamura2020} or measuring the operators $XX$ and $YY$ separately, but an advantage of this method is that Hamming weights of the final state can also be checked, allowing for a simple error detection procedure~\cite{Cade2020}.

Due to the application of $M$, two different hopping terms that have the qubit $i$ or $j$ in common cannot be measured at the same time. However, $M$ has the property that $M^\dagger (Z \otimes Z) M = Z \otimes Z$. In consequence, if the qubits are ordered $ijkl$ then the ``non-crossing'' hopping terms between $i - l$ and $j - k$ can be measured simultaneously, but hopping terms $i - k$ and $j - l$ cannot. This was used in~\cite{Cade2020} to measure all hopping terms in the 2D Hubbard model in 4 measurement rounds.

As before, by taking into account the structure of $H_\text{emb}$ we can reduce the number of circuit preparations required. Restricting to one spin type, it is possible to measure the hopping terms of $H_\text{emb}$ for the 1D Hubbard model in $N_\text{frag} + 1$ circuit preparations. This is the minimum bound possible since the two fragment edge sites each interact with $N_\text{frag} + 1$ other sites.

\begin{figure}[tb]
    \centering
    \includegraphics[width=\linewidth]{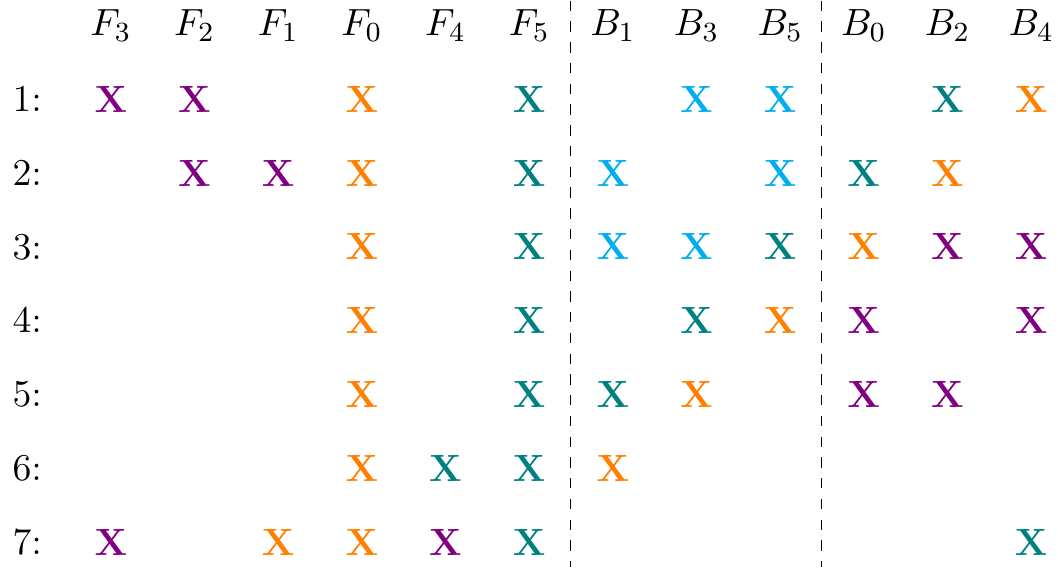}
    \caption{Demonstration of the measurement pattern for the 1D model using a fragment of size 6. The numbered lines are different preparations of the circuit. On each line, Xs of the same colour represent a hopping term that has been measured, with the different colours showing which terms can be measured simultaneously. Note that $M$ is applied on Xs of the same colour and then computational basis measurements are done on all the qubits.} 
    \label{fig:1Dmeasurement}
\end{figure}

Assuming the ordering in equation~(\ref{eq:1d_jw_ordering}), in the first circuit preparation we measure the terms $F_0 - B_{N_e}$ and $F_{N_\text{frag}-1} - B_{N_e-2}$. At subsequent preparations, we measure the fragment-bath terms by working our way down the bath sites, i.e.\ at the second preparation we measure $F_0 - B_{N_e-2}$ and $F_{N_\text{frag}-1} - B_{N_e-4}$. It is clear to see that in this way all of the hopping terms on the fragment edge sites can be measured in $N_\text{frag}+1$ runs of the circuit. The measurement of the remaining fragment- and bath-only terms can fit within these runs -- considering the measurement of the odd bath hopping terms using the general ``rainbow'' procedure described in Appendix~\ref{app:measurement}, this takes $\floor{N_\text{frag}/2}$ preparations and can be completed before the term $F_{N_\text{frag}-1} - B_{N_o}$ is measured. This is best demonstrated in Figure~\ref{fig:1Dmeasurement} with fragment size 6 as an example. 

The number of circuit preparations required to measure all of the terms for the 2D model is given in Table~\ref{tab:theory_layers}. We assume that the fragment- and bath-only hopping terms can be measured within the runs required for the fragment-bath terms. The fragment-bath terms can be measured in $N_\text{frag} + N_E - 1$ preparations where $N_E$ is the number of sites on the edge of the fragment. This is according to the procedure described in Appendix~\ref{app:measurement} for measuring all hopping terms between two sets of qubits. 

Note that we do not consider swapping the qubit ordering whilst doing measurements~\cite{Wecker2015_2} -- in some cases this can lead to fewer circuit preparations being required\footnote{The hopping terms for the 2D model with the 1D fragment could be measured in $\ceil{3N_\text{frag}/2}$ preparations rather than $2N_\text{frag}-1$ if we could swap the bath sites into different positions with every circuit preparation.} but leads to a higher gate depth. Another possibility for reducing the number of preparations is to change the Jordan-Wigner ordering of the bath sites at different circuit preparations~\cite{Cai2020}. We have also not considered this since it would change the order that terms are implemented in the ansatz. Numerical validation would be required to check that the ansatz performs in the same way.

\section{Numerical results}
\label{sec:results}

We ran numerical simulations up to a fragment size of 4 (16 qubits) for both the 1D and 2D Hubbard models using exact values for the expectation $\bra{\psi(\bm{\theta})} H_\text{emb} \ket{\psi(\bm{\theta})}$. We ran simulations up to a fragment size of 2 using measurements to estimate the expectation which will be discussed in Section~\ref{sec:results_measurements}. The code was written in C++ and the Quantum Exact Simulation Toolkit (QuEST)~\cite{quest} was used to simulate the quantum circuits. 

We ran two variants of the HV ansatz -- one with a low number of parameters per depth (equal to the minimal number of sets of commuting terms in $H_\text{emb}$), and one with a high number of parameters per depth (equal to the number of terms in $H_\text{emb}$, but affixing the same parameters to identical up and down terms). We call these ans\"atze HV-min and HV-max. The number of parameters per ansatz depth for HV-min is $O(N_\text{frag})$ and for HV-max is $O(N_\text{frag}^2)$. Taking 1D fragment shapes as an example, HV-min requires $N_\text{frag} + N_E + 1$ parameters per depth. On the other hand, HV-max requires $4N_\text{frag} + N_E N_\text{frag} + I(\ceil{N_\text{frag}/2}) + I(\floor{N_\text{frag}/2}) - 1$ parameters where $I(n) = n(n-1)/2$ is the maximum number of hopping interactions between $n$ qubits. 

Simulations were run up to an ansatz depth of 10 and 5 for HV-min and -max respectively. At each layer of the ansatz, we ran the onsite gates followed by the hopping gates and then number gates. Gates were implemented in the order that they would be in the swap network, including the reversal of the circuit at every other depth to fairly represent the behaviour of the actual quantum circuit that would be implemented on hardware. 

The Limited-memory Broyden-Fletcher-Goldfarb-Shanno (L-BFGS) optimisation algorithm provided by the nonlinear optimisation library NLopt~\cite{nlopt} was used for the VQE classical optimiser as we had found it to be effective in previous work~\cite{Cade2020}. We used the very simple secant method as a root-finding algorithm for $\mu$ in the DMET loop as it often converged within a few iterations without requiring gradient information. We set the secant method to terminate when within 0.1 of the root.

The infinite 1D and 2D Hubbard models were approximated with a 240 site and $20 \times 24$ site finite size model with anti-periodic boundary conditions. Simulations were run with fragment sizes of 1, 2, 3 and 4. In the 2D case with fragment size 4 we took the fragment shape to be $2 \times 2$ and did not consider $1 \times 4$. Recall that a fragment size of $N_\text{frag}$ requires $4N_\text{frag}$ qubits, so we simulated systems containing up to 16 qubits.

For each fragment size we fixed $t=1$ and initially ran experiments for $U = 1, 2, 4$ and 8 with quarter- and half-filling, which correspond to a fermion site occupancy of $\mean{n} = 0.5$ and 1. For each experiment using VQE as the DMET solver we ran a corresponding one using exact diagonalisation to compare the two solvers.

Tables~\ref{tab:depth1D} and~\ref{tab:depth2D} show the ansatz depths required to reach a 1\% relative error with the energy per site when using exact diagonalisation as the solver. The calculation of the energy per site at the end of the DMET algorithm is stated in Section~\ref{sec:observables}. The HV-max ansatz requires less depth to reach the same error than HV-min. However, this comes at a cost with the classical optimiser needing extra circuit evaluations. For $N_\text{frag}=2$, the optimiser took 5-10$\times$ more evaluations for the same ansatz depth. For the larger fragment sizes this went up to 10-20$\times$ with some extreme cases requiring up to $100\times$ more circuit runs.

\begin{figure*}[tb]
    \centering
    \includegraphics[width=0.9\linewidth]{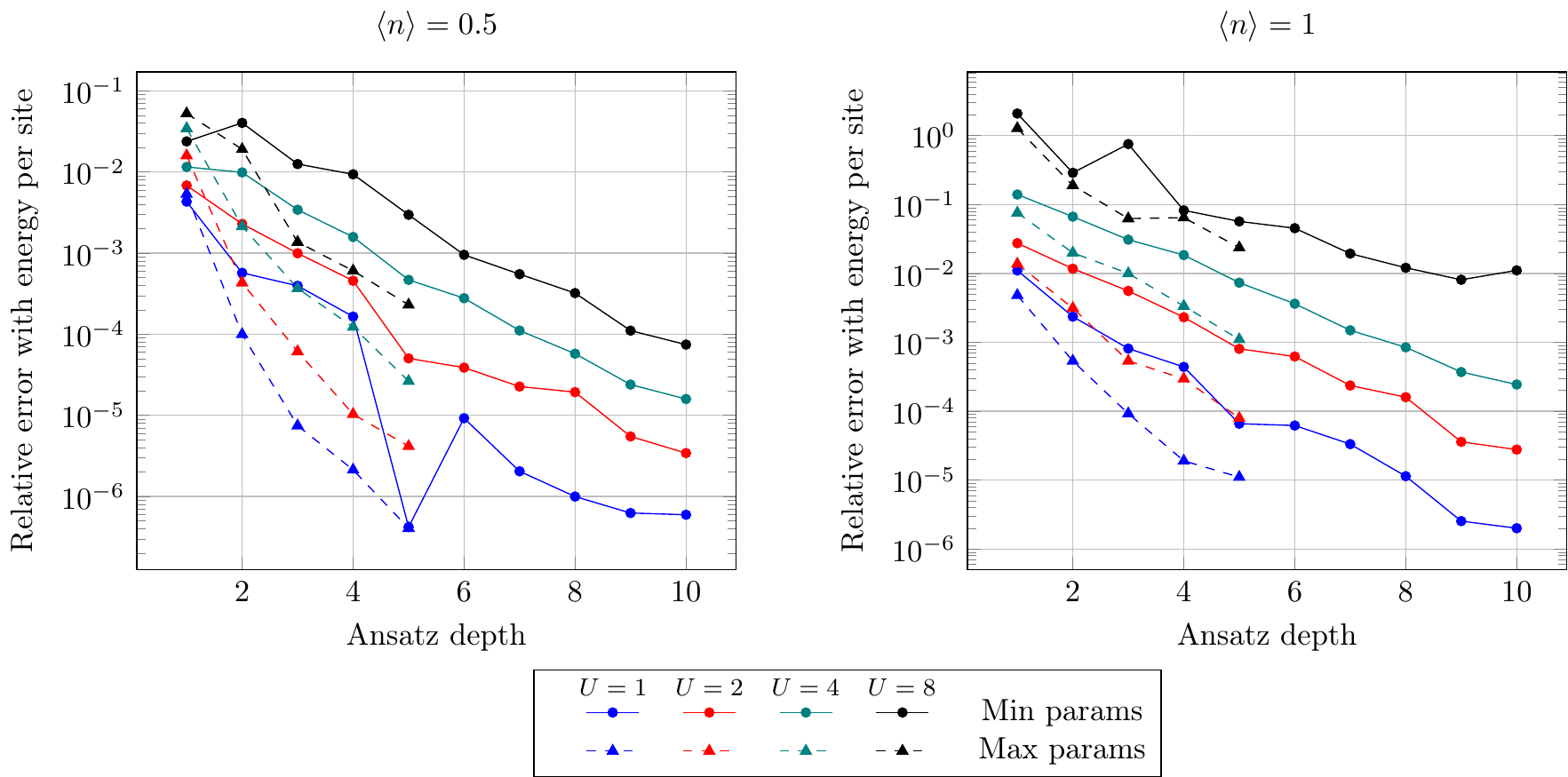}
    \caption{Comparison of HV-min and -max ans\"atze for the 2D model with the largest fragment size of $2 \times 2$ for quarter- and half-filling. The relative error is against the ground energy per site calculated with exact diagonalisation as the DMET solver.}
    \label{fig:error_vs_depth}
\end{figure*}

Note that the tables do not include fragment size 1 as the depths required to reach 1\% error were the same for both the 1D and 2D models and both variants of the HV ansatz. Depth 1 was required for $U=1, 2$ and $U=4$ with $\mean{n}=0.5$, and depth 2 for $U=4$ with $\mean{n}=1$ and $U=8$. The number of parameters required for the HV-min and -max ans\"atze were 3 and 4 respectively. Due to its small size, the behaviour of the ansatz as the depth increases was different from the larger fragment sizes. At depth 1, the error was typically on the order of $10^{-1} - 10^{-3}$ (depending on the value of $U$ and $\mean{n}$). This error dropped to $10^{-6} - 10^{-8}$ for depth 2 and plateaued for the other depths, meaning there is no benefit to going beyond depth 2 for the HV ansatz with $N_\text{frag}=1$. This is not the case for the other fragment sizes as increasing the depth almost always resulted in a lower error -- an example of this can be seen in Figure~\ref{fig:error_vs_depth} for a fragment size of $2 \times 2$.

\begin{table}[tb]
    \centering
\begin{tabularx}{\linewidth}{cc *{6}{Y}}
\toprule
 & 
 & \multicolumn{3}{c}{$N_\text{frag}$ (Min params)}  
 & \multicolumn{3}{c}{$N_\text{frag}$ (Max params)}\\
\cmidrule(lr){3-5} \cmidrule(l){6-8}
  $U$ & $\mean{n}$ & 2 (5) & 3 (6) & 4 (7) & 2 (11) & 3 (18) & 4 (25)\\
\midrule
 \multicolumn{1}{c}{\multirow{2}{*}{1}} & 0.5 & 1 & 1 & 1 & 2 & 1 & 1\\
 \multicolumn{1}{c}{}  & 1 & 2 & 1 & 1 & 2 & 1 & 1\\
 \midrule
 \multicolumn{1}{c}{\multirow{2}{*}{2}}  & 0.5 & 2 & 4 & 2 & 2 & 2 & 2\\
 \multicolumn{1}{c}{}  & 1 & 3 & 5 & 6 & 2 & 2 & 3 \\
 \midrule
 \multicolumn{1}{c}{\multirow{2}{*}{4}}  & 0.5 & 4 & 5 & 5 & 2 & 3 & 3 \\
 \multicolumn{1}{c}{}  & 1 & 4 & 8 & >10 & 3 & 4 & >5\\
  \midrule
 \multicolumn{1}{c}{\multirow{2}{*}{8}}  & 0.5 & 5 & 6 & 8 & 2 & 3 & 4\\
 \multicolumn{1}{c}{}  & 1 & 7 & >10 & >10 & 3 & 5 & >5\\
\bottomrule
\end{tabularx}
\caption{Depth of the ansatz required to achieve 1\% relative error against the ground energy per site, calculated with exact diagonalisation as the DMET solver for the 1D model using the HV ansatz with minimum and maximum number of parameters. The number of parameters is shown in brackets.}
\label{tab:depth1D}
\end{table}

\begin{table}[tb]
    \centering
\begin{tabularx}{\linewidth}{cc *{6}{Y}}
\toprule
 & 
 & \multicolumn{3}{c}{$N_\text{frag}$ (Min params)}  
 & \multicolumn{3}{c}{$N_\text{frag}$ (Max params)}\\
\cmidrule(lr){3-5} \cmidrule(l){6-8}
  $U$ & $\mean{n}$ & 2 (5) & 3 (7) & $2 \times 2$ (8) & 2 (11) & 3 (20) & $2 \times 2$ (32)\\
\midrule
 \multicolumn{1}{c}{\multirow{2}{*}{1}} & 0.5 & 1 & 1 & 1 & 1 & 1 & 1\\
 \multicolumn{1}{c}{}  & 1 & 2 & 1 & 2 & 1 & 1 & 1\\
 \midrule
 \multicolumn{1}{c}{\multirow{2}{*}{2}}  & 0.5 & 2 & 1 & 1 & 2 & 2 & 2\\
 \multicolumn{1}{c}{}  & 1 & 2 & 1 & 3 & 2 & 1 & 2\\
 \midrule
 \multicolumn{1}{c}{\multirow{2}{*}{4}}  & 0.5 & 2 & 1 & 2 & 2 & 2 & 2\\
 \multicolumn{1}{c}{}  & 1 & 3 & 4 & 5 & 2 & 2 & 4\\
  \midrule
 \multicolumn{1}{c}{\multirow{2}{*}{8}}  & 0.5 & 3 & 1 & 4 & 2 & 2 & 3\\
 \multicolumn{1}{c}{}  & 1 & 5 & 7 & 9 & 3 & 5 & >5\\
\bottomrule
\end{tabularx}
\caption{Depth of the ansatz required to achieve 1\% relative error with the ground energy for the 2D model.}
\label{tab:depth2D}
\end{table}

We found that the depth required increased as $U$ increased, which is to be expected as the starting state for the HV ansatz is the ground state for the $U=0$ embedded Hamiltonian. This can also be seen in Figure~\ref{fig:error_vs_depth} for solving the 2D model with a $2 \times 2$ fragment. The features of these two graphs -- that depth required increases as $U$ increases, that the depth required is higher for half-filling than quarter-filling and that HV-max requires roughly 2-3 fewer layers to get to the same accuracy as HV-min -- are representative of all the fragment sizes larger than 1. The VQE algorithm is a nested optimisation loop providing imperfect solutions to $H_\text{emb}$ within the larger DMET optimisation loop. The fact that the error in the energy per site goes down exponentially with the number of ansatz layers is an encouraging sign that the combination of DMET and VQE is effective. 

\begin{figure}[tb]
    \centering
    \includegraphics[width=0.9\linewidth]{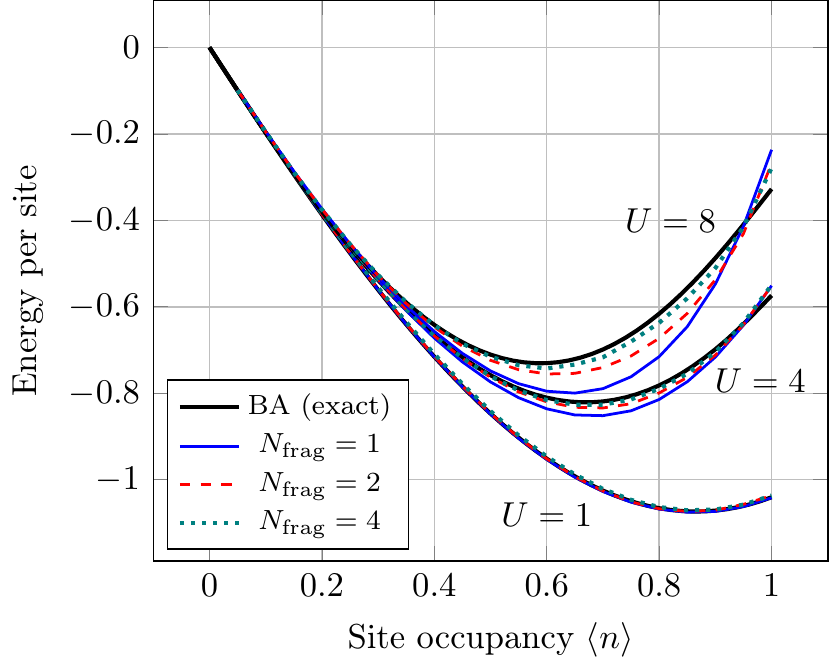}
    \caption{Plot of energy per site against site occupancy for the 1D model. The black line is the exact solution to the 1D Hubbard model calculated using the Bethe ansatz. The coloured lines are the values found using VQE as the DMET solver with 20 points taken between $\mean{n} = 0.05$ and 1. The ansatz used was HV-min. For $N_\text{frag} = 1$, the VQE depth used was 2, and for the other fragment sizes for each line we used the depths required for the given $U$ at half-filling in Table~\ref{tab:depth1D} (or depth 10 in the cases where 1\% relative error was not reached).}
    \label{fig:energy_curves}
\end{figure}

\begin{figure}[tb]
    \centering
    \includegraphics[width=0.9\linewidth]{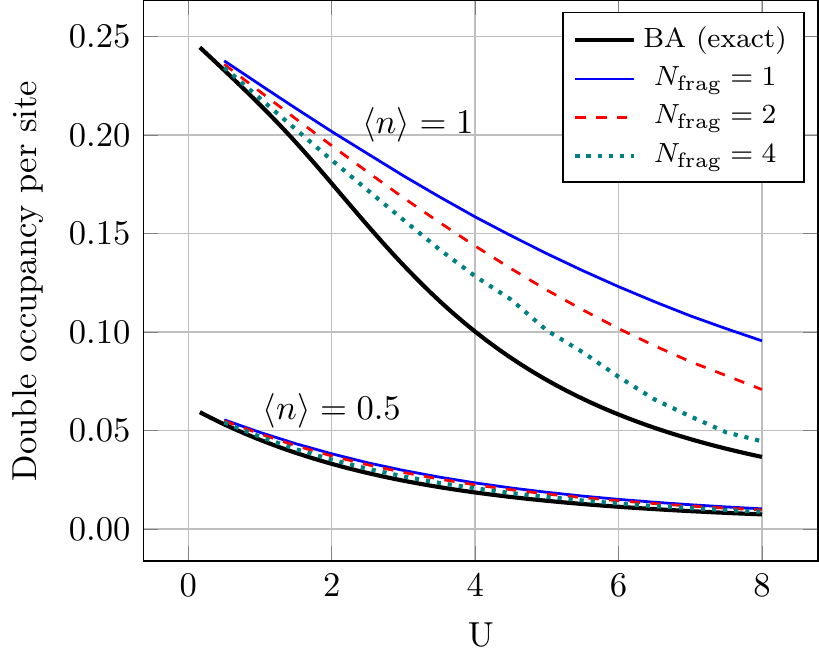}
    \caption{Plot of double occupancy per site against $U$ for the 1D model. The coloured lines are the values found using VQE as the DMET solver with 16 points taken between $U = 0.5$ and 8. The ansatz used was HV-min. For $N_\text{frag} = 1$, the VQE depth used was 2, and for the other fragment sizes we used the depths required for 1\% relative error with the double occupancy at $U=8$, which was generally the same as the depths in Table~\ref{tab:depth1D} except for $N_\text{frag}=4, \mean{n}=0.5$ where it was depth 10.}
    \label{fig:double_occ_curves}
\end{figure}

After running the batch of experiments discussed above, we carried out further experiments to compute physical properties of the Hubbard model. Figure~\ref{fig:energy_curves} is a plot of energy per site against site occupancy for the 1D model for $U=1,4,8$ and $N_\text{frag}=1,2,4$. The VQE ansatz used was HV-min and the depths chosen for the graph were those required for each $U$ at half-filling to reach 1\% error (see Table~\ref{tab:depth1D}). The 1D Hubbard model is exactly solvable using the Bethe ansatz~\cite{Shiba1972, Lieb1968} and has been plotted as a reference. The lines reproduce the behaviour seen using DMET in the original paper from Knizia and Chan~\cite{Knizia2012}.

A plot of double occupancy per site against $U$ is shown in Figure~\ref{fig:double_occ_curves} for the 1D model for quarter- and half-filling and $N_\text{frag}=1,2,4$. In the case of half-filling, the double occupancy curve is not reproduced well even when using a fragment size of 4. However, this deficiency is also present when using exact diagonalisation as the single-shot embedding solver, and is not a consequence of using VQE.

\subsection{Incorporating realistic measurements}
\label{sec:results_measurements}

We have shown that the VQE algorithm performs well as the solver for DMET when exact values are taken for the expectation values. Using exact values is a good test bed for trying out different ans\"{a}tze and checking if the algorithm can work in principle, but we also need to consider the more practical aspects of quantum computers such as measurements and noise. Here we run more experiments but include sampling from the quantum computer in the simulation. We do not consider any type of noise, hence our simulations represent an ideal quantum computer.  

Repeated measurement of states were simulated by storing the probability amplitudes of the state vector and then sampling from that discrete distribution. We picked a few representative simulations to re-run and used the simultaneous perturbation stochastic approximation (SPSA) optimisation algorithm~\cite{spsa, Spall1998} in place of L-BFGS. SPSA is a form of stochastic gradient descent where a gradient is taken in one random direction (instead of all directions); it is designed to be robust to noise and require fewer function evaluations. SPSA has been shown to be effective with VQE~\cite{Cade2020, Sung2020} and has been used in experiments on quantum hardware~\cite{Kandala2017, Ganzhorn2019, Montanaro2020}.

We picked the $U=4$ 1D Hubbard model and ran the HV-min ansatz up to a fragment size of 2 with a range of fillings. The SPSA meta-parameters~\cite{spsa} were set to be $\alpha=0.602, \gamma=0.101$ (the theoretically optimal values~\cite{Spall1998}); $c=0.2$ (from~\cite{Cade2020}); and $a=2, A=10$ (to allow for fast convergence). Each term in the expectation $\mean{H_\text{emb}}$ was estimated using $10^4$ samples and the final state at the end of the SPSA algorithm with $10^5$ samples. The maximum number of SPSA iterations was set to be 2,000 for $N_\text{frag}=1$ and $10,000$ for $N_\text{frag}=2$. As before we use the secant method to find $\mu$ in the DMET optimisation loop but loosen the termination criteria to stop if within 0.5 of the root.

\begin{figure}[tb]
    \centering
    \includegraphics[width=0.9\linewidth]{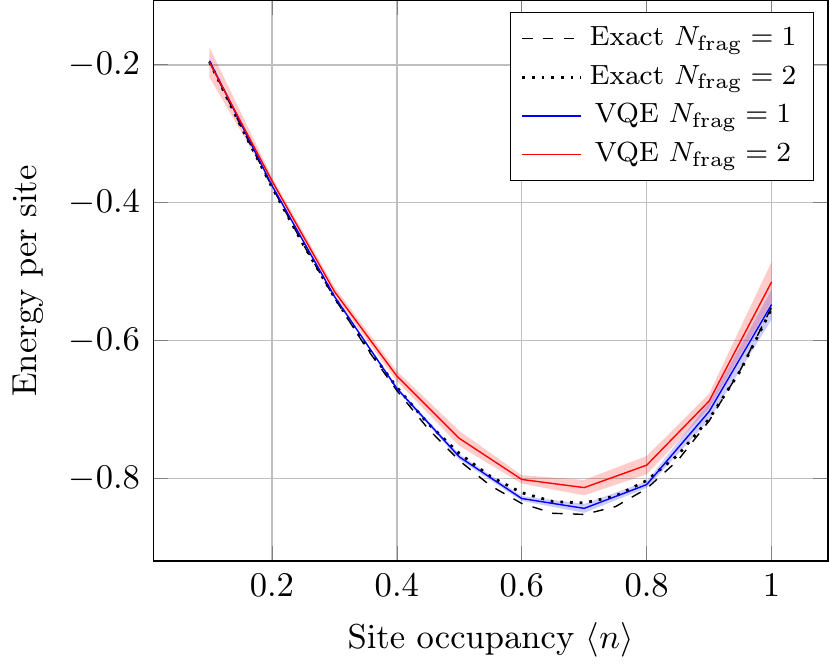}
    \caption{Plot of energy per site for the $U=4$ 1D Hubbard model solved with fragment sizes 1 and 2 using exact diagonalisation and VQE (with measurements) as the solver. The ansatz circuit is HV-min and the VQE depth is 2 for $N_\text{frag}=1$ and 4 for $N_\text{frag}=2$. 10 points were taken between $\mean{n}=0.1$ and 1. The solid lines show the mean of 10 DMET runs, with the shaded region being the standard deviation.}
    \label{fig:energy_curve_measurements}
\end{figure}

Figure~\ref{fig:energy_curve_measurements} is a plot of the energy per site against site occupancy when using VQE with sampling as the DMET solver. Solving fragment size 1 $H_\text{emb}$ with VQE reproduces the exact diagonalisation results with on average 0.5-1.5$\%$ relative error. However the fragment size 2 VQE curve has a larger relative error of around 2-5$\%$. The fidelities\footnote{The fidelity of two quantum states $\ket{\phi}$ and $\ket{\psi}$ is defined to be $F = |\mean{\phi | \psi}|^2$.} of the ground states output from SPSA with the ground state of $H_\text{emb}$ were typically above 0.999 for $N_\text{frag}=1$ and around 0.985-0.995 for $N_\text{frag}=2$. It is likely that by changing the SPSA meta-parameters or by using a different optimisation method, the fidelity for fragment size 2 could be increased.

\begin{figure}[tb]
    \centering
    \includegraphics[width=0.9\linewidth]{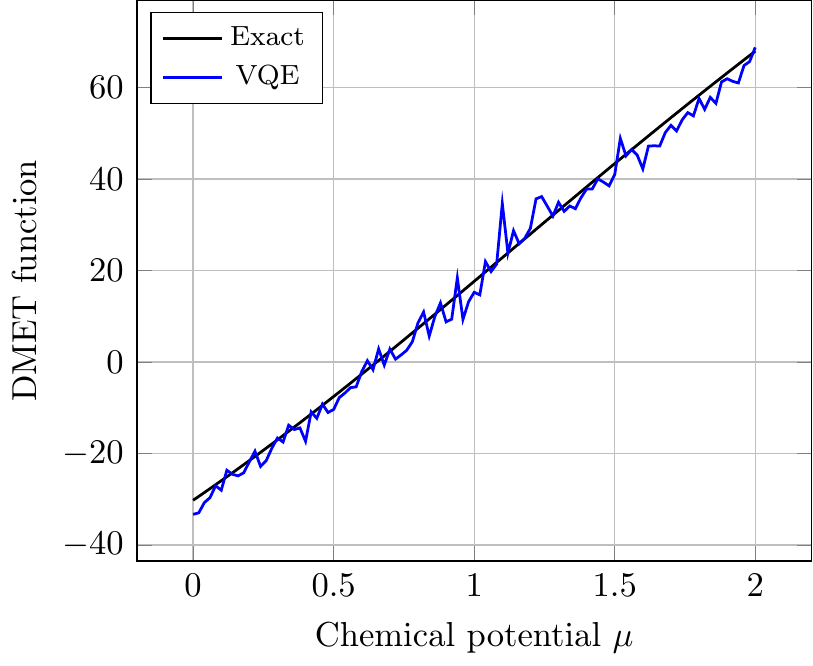}
    \caption{Sweeping through the chemical potential for 100 values between $\mu=0$ and 2. At each value of $\mu$, the embedded Hamiltonian is solved and the DMET function is calculated. The graph shows the $U=4$ 1D Hubbard model with $\mean{n}=0.5 \,(N_\text{occ}=120)$ and a fragment size of 1. The VQE solver uses depth 2 HV-min as the ansatz circuit. Note that although the DMET function is a straight line here, this is not always the case.}
    \label{fig:mu_sweep}
\end{figure}

In addition to the problem of whether the optimisation method chosen for VQE will converge, we must also consider whether the root-finding technique used to find $\mu$ will be able to handle the extra statistical noise. Figure~\ref{fig:mu_sweep} demonstrates what happens to the DMET function (see Appendix~\ref{app:single_shot_embedding}) when using VQE with sampling as the solver. 

Despite the secant method having to deal with a noisier function, we found that in practice it still performed well most of the time and usually converged in less than 10 iterations. As a test for this we bypassed the secant method by solving $H_\text{emb}$ with the optimal value of $\mu$ using SPSA and found that the plot of the energy per site was similar to Figure~\ref{fig:energy_curve_measurements}. 

Occasionally we found that the secant method became unstable and did not converge. This happened quite rarely (for example for one filling and one run out of the 10 runs) and so we re-ran the simulations where this occurred. When running on quantum hardware this could be dealt with by monitoring which values of $\mu$ the secant method picks or by averaging multiple runs of SPSA for a certain $\mu$. Other possibilities include combining the secant method with a curve fitting technique so that all of the known information about the DMET function can be effectively used, or squaring the function and using a gradient descent algorithm to find the minimum.

\section{Conclusion}

We have carried out a detailed study into how single-shot DMET could be used to solve the Hubbard model on a quantum computer with VQE. We have used the form of the embedded Hamiltonian to construct efficient swap networks for implementing the HV ansatz, and measurement schemes for estimating expectation values. These constructions have assumed that we are using the Jordan-Wigner encoding and the architecture of the quantum device is fully-connected. 

We also conducted numerical simulations up to a fragment size of 4 (16 qubits) using exact expectation values from the VQE, and up to fragment size 2 (8 qubits) involving measurements. These are the largest simulations done to date for the combination of DMET and VQE. The VQE algorithm is a nested optimisation loop providing imperfect solutions to the embedded Hamiltonian within the larger DMET optimisation loop. There is a lot of scope for errors to propagate throughout the algorithm, but despite this the simulations showed that DMET with VQE was effective. The errors on the observables were shown to decrease exponentially with the depth of the ansatz, meaning that it is possible to use a lower depth ansatz if a high accuracy is not required.

DMET is an embedding method which can be used to drastically reduce the number of qubits required to find the ground state properties of a given Hamiltonian. However, in the case of the Hubbard model, applying the embedding procedure leads to an embedded Hamiltonian with a higher complexity than the original one.

This suggests that DMET may be most appropriate in the regime where quantum hardware has a low number of qubits that are relatively noise free, allowing circuits of high depth to be implemented. Complicated molecules can require thousands of qubits to simulate; by using DMET the qubit count could be greatly reduced and using a quantum computer could allow access to larger fragment sizes than is currently possible. There have already been several small-scale demonstrations of DMET on quantum hardware~\cite{Tilly2021, Kawashima2021} but more research is needed into the effect of measurements and noise on the DMET algorithm. \\

Data are available at the University of Bristol data repository, data.bris~\cite{data}.

\subsection*{Acknowledgements}

We would like to thank the Phasecraft team for useful discussions and support throughout this project. L.M. would also like to thank John R. Scott for helpful discussions whilst working from home. L.M. received funding from the Bristol Quantum Engineering Centre for Doctoral Training, EPSRC Grant No. EP/L015730/1. This project has received funding from the European Research Council (ERC) under the European Union's Horizon 2020 research and innovation programme (grant agreement No.\ 817581). Google Cloud credits were provided by Google via the EPSRC Prosperity Partnership in Quantum Software for Modeling and Simulation (EP/S005021/1).

\appendix
\section{The single-shot embedding algorithm} 
\label{app:single_shot_embedding}

Here we lay out the steps of the single-shot embedding algorithm given in Section~\ref{sec:single_shot_embedding} in greater detail. The algorithm presented here is from~\cite{Wouters2016} where the steps of a general DMET calculation are also explained. Recall that we are reducing the problem of solving an $N$ site Hubbard model occupied by $N_\text{occ}$ fermions to an embedded problem with $N_\text{frag}$ sites in the fragment. 

\begin{enumerate}
    \item \textit{Calculate the ground state of the approximating mean-field Hamiltonian}\\
    The simplest form the mean-field Hamiltonian can take is the one-particle (quadratic) part of $H_\text{hub}$, 
    \begin{equation}
        H_{MF} = T = -t \sum_{\mean{i, j}, \sigma} a^\dagger_{i\sigma} a_{j\sigma}.
    \end{equation}
    Note that $H_{MF}$ could be chosen to be different from $T$, which would lead to a different embedded Hamiltonian. We must find the 1-RDM of the ground state of $H_{MF}$. The 1-RDM expresses the relationship between the behaviour of an electron at two different sites and the diagonal contains electron densities. For a state $\ket{\psi}$ it is defined to be 
    \begin{equation}
        \rho(\ket{\psi})_{ij} = \bra{\psi} a^\dagger_j a_i \ket{\psi}.
    \end{equation}
    $H_{MF}$ is a quadratic Hamiltonian which can be solved efficiently and its ground state is a Slater determinant. This ground state can be found by taking the matrix $C$ of coefficients of $H_{MF}$. Restricting to one spin type since in this case both spins are identical, $C$ is an $N \times N$ matrix with the $(i, j)^\text{th}$ element being the coefficient of $a^\dagger_{i\uparrow} a_{j\uparrow}$ in the Hamiltonian. Without loss of generality we assume that the orbitals have been ordered such that the environment sites follow the fragment sites.
    
    We then diagonalise $C$ and put the eigenvectors corresponding to the lowest $N_\text{occ}/2$ eigenvalues (recall that half of the electrons are spin up) into an $N \times N_\text{occ}/2$ matrix $\Phi$ which now represents the ground state Slater determinant (each column is an occupied orbital written as a linear combination of the original orbitals). Since $\Phi$ is a Slater determinant, its 1-RDM can be simply calculated as
    \begin{equation}
        \rho(\Phi) = \Phi \Phi^\dagger.
    \end{equation}
    A derivation for this fact is provided in Appendix~\ref{app:derivations}.
    
    \item \textit{Construct the projector from the mean-field ground state}\\
    The 1-RDM of the ground state is used to construct the projector that will reduce the environment orbitals to the bath orbitals. Delete the first $N_\text{frag}$ rows and columns of the 1-RDM. We are left with an $N_\text{env} \times N_\text{env}$ sub-matrix (where $N_\text{env} = N - N_\text{frag}$) representing the environment orbitals. Diagonalising this submatrix leads to 3 different scenarios:
    \begin{itemize}
        \item Eigenvalues of 0 correspond to unoccupied environment orbitals.
        \item Eigenvalues of 1 correspond to occupied environment orbitals. Counting these tells us the occupation number of the embedded Hamiltonian we will need to solve in step 4. If there are $m$ of these then the embedded occupation number including both spin types is $N_\text{emb} = N_\text{occ} - 2m$.
        \item Eigenvalues between 0 and 1 have overlap on the environment and the fragment. There will be $N_\text{frag}$ of these and we will write the eigenvectors associated to them as $v_i$.
    \end{itemize}
    The eigenvectors associated to eigenvalues of 0 and 1 are discarded and the rest are used to define the projector
    \begin{equation} \label{eq:projector}
        P = \begin{pmatrix} 
            I & 0 \\
            0 & v_1...v_{N_\text{frag}}
        \end{pmatrix},
    \end{equation}
    where $I$ is the identity matrix of size $N_\text{frag} \times N_\text{frag}$. Note that this procedure outlined in steps 1 and 2 is equivalent to finding the Schmidt decomposition of $\ket{\Phi}$ to calculate the projector~\cite{Wouters2016}.
   
    \item \textit{Construct the embedded Hamiltonian from the projector} \\
    The embedded Hamiltonian is constructed using the non-interacting bath formulation~\cite{Wouters2016} where only the quadratic part of $H_\text{hub}$ is projected and higher order terms are only added back to the fragment. This is a simpler construction than projecting the full Hamiltonian $H_\text{hub}$ which can lead to more complicated interaction terms. 
    
    The projection of the quadratic part $T$ of $H_\text{hub}$ into the embedded basis is obtained as follows. Write $T = T^\uparrow + T^\downarrow$ and interpret each term as a matrix of coefficients $K^\uparrow$ and $K^\downarrow$, similarly to $C$ in step 1. Now project the matrices of coefficients into the embedded basis to obtain 
    \begin{equation}
        K^\sigma_{\text{emb}} = P^\dagger K^\sigma P, \quad \sigma \in \{\uparrow, \downarrow\}.
    \end{equation}
    The $K^\sigma_\text{emb}$ can then be re-interpreted as Hamiltonians $T^\sigma_\text{emb}$, which can be added up to obtain $T_\text{emb}$.

    Moving onto the two-particle interaction term in the embedded Hamiltonian, this is simply set to be the terms in $W$ that act only on the fragment,
    \begin{equation}
        W_\text{emb} = U \sum_{i \in \text{frag}} n_{i\uparrow} n_{i\downarrow}.
    \end{equation}
    Finally, a chemical potential term $\mu$ that governs the number of electrons in the fragment is also added to the embedded Hamiltonian. This is the only parameter that is determined self-consistently in this variant of DMET and makes the embedded Hamiltonian
    \begin{equation} 
        H_{\text{emb}} = T_\text{emb} + W_\text{emb} - \mu \sum_{i \in \text{frag}, \sigma} n_{i\sigma}.
    \end{equation}
    
    \item \textit{Solve the embedded problem} \\
    $H_\text{emb}$ is a Hamiltonian on $4N_\text{frag}$ orbitals ($2N_\text{frag}$ for each spin's fragment and bath sites). The ground state $\ket{\Phi_\text{emb}}$ of $H_\text{emb}$ occupied by $N_\text{emb}$ electrons can be found using methods such as exact diagonalisation, DMRG, or VQE.
    
    \item \textit{Adjust the chemical potential until there are the correct number of particles in the fragment} \\
    Repeat from step 3, adjusting $\mu$ until the fraction of occupied orbitals in the fragment matches the site occupancy of $H_\text{hub}$. Since $\mu$ is only one parameter, it can be fitted by finding roots of $f(\mu) = 0$ where
    \begin{equation}
        \qquad f(\mu) = \frac{N}{N_\text{frag}} \sum_{i \in \text{frag}, \sigma} \bra{\Phi_\text{emb}} n_{i\sigma} \ket{\Phi_\text{emb}} - N_\text{occ}.
    \end{equation}
    In the paper we refer to this as the DMET function. $f(\mu)$ is equal to the number of electrons in the fragment scaled up to fill the large model, minus the number of electrons in the Hubbard model to be solved for.
    
\end{enumerate}

In a general DMET calculation the system can be split into multiple disjoint fragments, with the bath for each fragment constructed from the union of the other fragments. Consistency then has to be enforced between all the separate fragment-bath systems~\cite{Knizia2013, Wouters2016}. We do not need to consider this as the Hubbard model is translationally invariant. This makes the use of multiple fragments redundant as they would all have the same properties.

\begin{figure}[tb]
    \centering
    \includegraphics[width=0.9\linewidth]{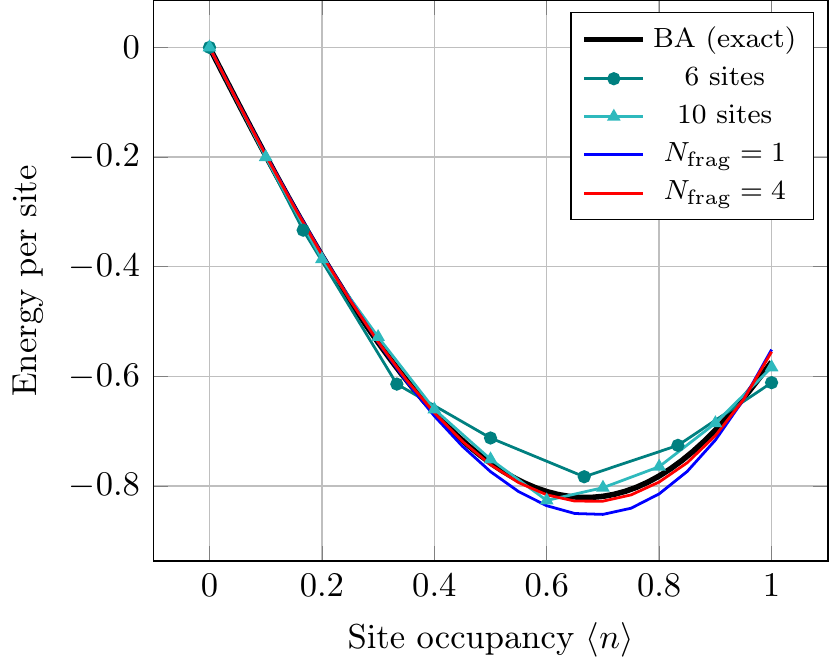}
    \caption{Energy per site for the 1D $U=4$ Hubbard model solved with the Bethe ansatz, exact diagonalisation for small models with periodic boundary conditions and single-shot embedding.}
    \label{fig:dmet}
\end{figure}

We conclude this appendix with a demonstration of the effectiveness of single-shot embedding for the Hubbard model. In Figure~\ref{fig:dmet} we estimate the energy per site for the infinite 1D Hubbard model using  
single-shot embedding with fragment sizes of 1 and 4 (4 and 16 qubits), and exact diagonalisation for small Hubbard models of 6 and 10 sites (12 and 20 qubits) with periodic boundary conditions. On a small quantum computer it may be preferable to run DMET as it can achieve a high accuracy with a small number of sites. It is also possible to input any fraction for $\mean{n}$ with DMET, whereas with small models the site occupancy will be restricted to values for which there are a whole number of electrons.

\section{Calculating the 1-RDM of a Slater determinant}
\label{app:derivations}

In this appendix we will show that the 1-RDM $\rho$ of a Slater determinant is $\rho = \Phi \Phi^\dagger$, where $\Phi$ is the matrix representation of a Slater determinant (each column is an occupied orbital written as a linear combination of the original orbitals). 

Let $N$ be the number of spin-orbitals in a system, $M$ of which are occupied by fermions. An arbitrary Slater determinant $\ket{\Psi}$ can be written as
\begin{equation} \label{eq:slater_det}
    \ket{\Psi} = \prod_{\mu=1}^M c_\mu^\dagger \ket{\text{vac}},
\end{equation}
where $\{c_\mu^\dagger \}_{\mu=1}^M$ are the occupied orbitals and $ \ket{\text{vac}}$ is the vacuum state. The occupied orbitals can be written in terms of the original spin-orbitals $\{a_k^\dagger \}_{k=1}^N$,
\begin{equation} \label{eq:occ_orbitals}
    c_\mu^\dagger = \sum_{k=1}^N a_k^\dagger \Phi_{k\mu}
\end{equation}
for $\mu = 1,\hdots,M$, where $\Phi$ is an $N \times M$ matrix of coefficients writing the occupied orbitals in terms of the original orbitals~\cite{Wouters2016}. The $\{a_k^\dagger \}$ form a basis for the system, but the $\{c_\mu^\dagger \}$ may not since there are only $M \leq N$ of them. However, they can be expanded to form a basis by adding in $N-M$ more linearly independent $c_\mu^\dagger$ so that equation~(\ref{eq:occ_orbitals}) now applies for $\mu = 1,\hdots,N$. A consequence of this is that the original orbitals can now be written in terms of the occupied orbitals as
\begin{equation}
    a_k^\dagger = \sum_{\mu=1}^N c_\mu^\dagger \Phi_{\mu k}^* \quad \text{and} \quad a_k = \sum_{\mu=1}^N c_\mu \Phi_{k\mu}.
\end{equation}

Now calculating the 1-RDM becomes
\begin{align*}
    \rho_{ij} &= \bra{\Psi} a_j^\dagger a_i \ket{\Psi} \\
    &= \bra{\Psi} \left(\sum_{\mu=1}^N c_\mu^\dagger \Phi_{\mu j}^* \right) \left(\sum_{\nu=1}^N c_\nu \Phi_{i\nu} \right) \ket{\Psi} \\
    &= \sum_{\mu\nu}^N \Phi_{i\nu} \Phi_{\mu j}^* \bra{\Psi} c_\mu^\dagger c_\nu \ket{\Psi} \\
    &= \sum_{\mu\nu}^M \Phi_{i\nu} \Phi_{\mu j}^* = \Phi \Phi^\dagger,
\end{align*}
since
\begin{equation}
    \bra{\Psi} c_\mu^\dagger c_\nu \ket{\Psi} = \begin{cases}
    1, & \text{if } \mu=\nu \text{ and } \mu=1,\hdots,M \\
    0, & \text{otherwise}
    \end{cases}
\end{equation}
from the definition of the Slater determinant in equation~(\ref{eq:slater_det}).

\section{Deriving the form of the embedded Hamiltonian}
\label{app:form_emb_hamiltonian}

In this Appendix we present a derivation for why the embedded Hamiltonian takes the form that it does when solving the 1D Hubbard model. In particular, we will explain why the bath hopping terms split into two groups where in the first group all the even numbered sites interact with each other, and in the second group odd sites interact. We will show that this occurs using periodic boundary conditions when $N_\text{occ}/2$ is odd, or with anti-periodic boundary conditions when $N_\text{occ}/2$ is even. We also briefly discuss the form of $H_\text{emb}$ for the 2D model and the conditions under which the bath sites split into four groups.

\subsection{1D Hubbard model}

We will be following through the first 3 steps of the single-shot embedding algorithm from Appendix~\ref{app:single_shot_embedding}. The first step requires us to find the 1-RDM $\rho$ of the ground state of $T$ which is done by considering $K^\uparrow$, the $N \times N$ coefficient matrix restricted to one spin type. For simplicity of notation, we will refer to this matrix as $T$ for the rest of this appendix. We will also take $t=1$ in $T$ and let $M = N_\text{occ}/2$.

Let $\Phi = (v_0, v_1, \hdots, v_{M-1})$ where $v_i$ are the eigenvectors of $T$ associated to the lowest $M$ eigenvalues. The 1-RDM can be written as
\begin{equation}
    \rho = \Phi \Phi^\dagger = \sum_{i=0}^{M-1} v_i v_i^\dagger = \sum_\lambda P_\lambda,
\end{equation}
where the $P_\lambda$ group together the $v_i v_i^\dagger$ where $v_i$ share the same eigenvalue $\lambda$. If the $v_i$ contained in $P_\lambda$ spans the full eigenspace, then $P_\lambda$ is a projector onto the full eigenspace of $\lambda$.

Let us first consider the case where $T$ is periodic, i.e. $T_{i, i+1}=T_{i+1, i}=T_{0, N-1} = T_{N-1, 0} = -1$ for $i=0,\hdots,N-2$ and $T_{ij}=0$ otherwise. $T$ is a circulant matrix which therefore commutes with the cyclic permutation matrix $S$ given by $S_{i+1,i}=S_{0,N-1}=1$ for $i=0,\hdots,N-2$ and $S_{ij}=0$ otherwise. Commuting matrices preserve each other's eigenspaces and in particular commute with the projectors onto each other's eigenspaces. This means that $S$ commutes with $P_\lambda$, and therefore $\rho$, provided that the $P_\lambda$ project onto whole eigenspaces. 

Whether the $P_\lambda$ project onto whole eigenspaces depends on $M$ if there are eigenvalues with multiplicities greater than 1. To determine this we need to find the eigenvalues of $T$, which turns out to be a simple task since the eigenvalues/vectors of circulant matrices are well known~\cite{Bamieh2018}. The eigenvalues of $T$ are given by $\lambda_j = -2\cos(2\pi j/N)$ for $j = 0,\hdots,N-1$. There is one eigenvalue of -2 and one of 2 (if $N$ is even), and the rest all come in pairs. As a consequence $M$ must be odd for the eigenvector pairs to be included in $\Phi$.

If $M$ is odd, $\rho$ commutes with $S$. A matrix is circulant if and only if it commutes with $S$~\cite{Bamieh2018}, therefore $\rho$ is also circulant. It is also trivial to show that it is symmetric from $\Phi \Phi^\dagger$. These two properties of the 1-RDM will be used when working through step 2 of the DMET algorithm, but before proceeding we can perform a similar analysis to find the structure of $\rho$ in the anti-periodic case.  

Let $T'$ be the matrix associated to the anti-periodic model, i.e. $T'_{i, i+1}=T'_{i+1, i}= -1$ for $i=0,\hdots,N-2$, $T'_{0, N-1} = T'_{N-1, 0} = 1$ and $T'_{ij}=0$ otherwise. $T'$ is almost circulant but a minus sign is introduced when an element wraps back to the first column of the matrix. We will refer to this as an almost-circulant matrix. $T'$ commutes with $S'$ where $S'_{i+1,i}=1$ for $i=0,\hdots,N-2$, $S'_{0,N-1}=-1$ and $S'_{ij}=0$ otherwise. 

Following a similar argument to before, $S'$ will commute with $\rho$ if the $P_\lambda$ project onto whole eigenspaces. The eigenvalues of $T'$ can be shown to be $\lambda'_j = -2\cos((2j+1)\pi/N)$ for $j = 0,\hdots,N-1$, which come in pairs. Therefore $M$ must be even for $\rho$ to commute with $S'$.

We must now determine what the structure of $\rho$ is using the fact that it commutes with $S'$. If we have ${\rho = S'^T \rho S'}$ then we can match the right and left hand sides to get $\rho_{i,j} = \rho_{i+1, j+1},\, \rho_{N-1, i}=-\rho_{0, i+1},\, \rho_{i, N-1}=-\rho_{i+1, 0}$ and $\rho_{N-1, N-1}=\rho_{00}$ for $i,j=0,\hdots,N-2$. This implies that $\rho$ is almost-circulant. 

We have now shown that when (anti-)periodic boundary conditions are combined with $N_\text{occ}/2$ (even)odd, then the 1-RDM is (almost-)circulant and symmetric. These matrices have another property, that they are both Toeplitz. This follows from the (almost-)circulant property, but some intuition for this is that the Hubbard model is translationally invariant, therefore its 1-RDM should depend only on the distance between sites. Symmetric Toeplitz matrices are well studied and have nice properties that will allow us to find the form of the projector in step 2 of the DMET algorithm.

To calculate the projector $P$ onto the embedded basis, we take the submatrix $\rho_E$ of $\rho$ that corresponds to the environment and calculate its eigenvalues/vectors. $\rho_E$ has $N_\text{frag}$ eigenvalues between 0 and 1 and the rest are either exactly 0 or 1~\cite{Wouters2016}. Since $\rho_E$ is symmetric Toeplitz, its eigenvectors will split as evenly as possible into symmetric and skew-symmetric\footnote{Let $J$ be the $N\times N$ matrix $J_{ij}=\delta_{i, N-1-j}$. A vector $v$ is symmetric if $Jv=v$ and skew-symmetric if $Jv=-v$.}. This equal split of eigenvectors applies to the eigenspaces as well~\cite{Delsarte1983}. Therefore the eigenspaces associated to the eigenvalues of 0 and 1 will split as evenly as possible into symmetric and skew-symmetric, leaving an equal split of eigenvectors for the eigenvalues between 0 and 1 as well. This means that when the eigenvectors corresponding to the eigenvalues between 0 and 1 are placed in $P$ according to equation (\ref{eq:projector}), half of them will be symmetric and half skew-symmetric.

We can now move onto step 3 of the single-shot embedding algorithm and project $T$ using $P$ to obtain 
\begin{align}
    T_\text{emb} &= P^\dagger T P \nonumber\\ &=
    \begin{pmatrix}
    I & 0 \\
    0 & V^T
    \end{pmatrix} 
    \begin{pmatrix}
    A_F & B \\
    B^T & A_E
    \end{pmatrix} 
    \begin{pmatrix}
    I & 0 \\
    0 & V 
    \end{pmatrix} \nonumber\\
    &= 
    \begin{pmatrix}
    A_F & BV \\
    (BV)^T & V^T A_E V
    \end{pmatrix}, 
    \label{eq:T_emb}
\end{align}
where the definitions of the submatrices are as follows. $I$ is the identity matrix of size $N_\text{frag}$ and $0$ is the matrix of zeros. $V$ is the matrix of eigenvectors of $\rho_E$ that were placed in $P$, it is of size $N_\text{env} \times N_\text{frag}$ and since $\rho_E$ is symmetric, $V$ is real. $A_{F/E}$ are the matrices with $-1$s on the off-diagonal and are of size $N_\text{frag} \times N_\text{frag}$ and $N_\text{env} \times N_\text{env}$ respectively. $B$ has a $-1$ in the bottom-left corner and $\pm 1$ in the top-right (depending on which boundary conditions are used for $T$).

It can be seen from equation~(\ref{eq:T_emb}) that $A_F$ defines the fragment-only interactions, $BV$ the fragment-bath and $V^T A_E V$ the bath-only interactions. The hopping terms on the fragment have been preserved and are nearest-neighbour. Due to the structure of $B$, $BV$ has zeros everywhere except the top and bottom row. This means that the fragment sites on the ends of the fragment interacting with every bath site. 

Finally, we turn to the bath-only interactions. $A_E$ preserves the space of symmetric and skew-symmetric vectors, therefore the columns of $A_E V$ are all symmetric or skew-symmetric. When the matrix multiplication $V^T A_E V$ is done, the symmetric rows of $V^T$ will cancel with the skew-symmetric columns of $A_E V$ (and vice-versa), leading to zeroes in the bath part of $T_\text{emb}$. This corresponds to the bath sites splitting into two equal sized groups, where inside each group all of the sites share hopping terms. If we did not know that $V$ contained symmetric and skew-symmetric eigenvectors then we could not show that the bath-only hopping terms have this structure, and $V^T A_E V$ could be completely dense.

We have observed in practice that when the eigenvectors in $V$ are ordered according to their eigenvalues, they alternate symmetric and skew-symmetric. This is where the split into even and odd bath sites comes in. This property is called interleaving and in general is hard to prove. For our purposes, it is sufficient to show the split into equally sized groups for our purpose.

This analysis can be applied to any mean-field Hamiltonian that has a matrix of coefficients that is circulant or almost-circulant. The multiplicities of the eigenvalues of the Hamiltonian could lead to different restrictions on $N_\text{occ}$. In addition, the types of fragment-bath interactions that occur could be different, as this will depend on the form of $B$. For example, for the Hubbard model with next-nearest neighbour interactions, the two fragment sites closest to each end will interact with all of the bath sites.

\subsection{2D Hubbard model}

Let $T_n$ be the coefficient matrix of hopping terms for the $n$ site 1D Hubbard model with (anti-)periodic boundary conditions. The hopping matrix $T$ for the 2D $n \times m$ model is $T = T_n \otimes I_m + I_n \otimes T_m$, where $I_n$ is the identity matrix of size $n$. If the eigenvalues of $T_n$ are given by $\lambda_{ni}$ and the eigenvectors by $v_{ni}$, then the eigenvalues of $T$ are $\lambda_{ij} = \lambda_{ni} + \lambda_{mj}$ and the eigenvectors $v_{ij} = v_{ni} \otimes v_{mj}$. 

Taking periodic boundary conditions as an example, it is clear to see that $T$ commutes with $S_n \otimes S_m$ where $S_n$ is the cyclic permutation matrix of size $n \times n$.  Similar to the 1D case, we find that if whole eigenspaces are included in $\Phi$ then the 1-RDM also commutes with $S_n \otimes S_m$ and it turns out to be block circulant with circulant blocks (with the block sizes depending on $n$ and $m$).

However, this structure does not remain in the submatrix $\rho_E$, making an analysis like the previous one very difficult. We observed that when the fragment shape is 1D, the bath sites split into odd and even groups. When the shape of the fragment is 2D, the bath sites split into four roughly equal groups and the grouping of terms changes with different input parameters to the model. These splits only happen when full eigenspaces are included in $\Phi$. Since the multiplicities of the eigenvalues $\lambda_{ij}$ depend on $n$ and $m$, so do the allowed values of $N_\text{occ}$.

\section{Swap network for the 2D model}
\label{app:2d_swap_network}

When $H_\text{hub}$ is 2D but the shape of the fragment is 1D, recall from Section~\ref{sec:form_emb_hamiltonian} that the structure of $H_\text{emb}$ is similar to that of the 1D model except now all of the fragment sites interact with all of the bath sites. $2N_\text{frag}~-~1$ layers of two-qubit gates are required to implement all the necessary hopping terms. Consider two sets $A$ and $B$ of $N_A$ and $N_B$ qubits. If every qubit in set $A$ has a hopping term with every qubit in $B$ and the Jordan-Wigner ordering has all the qubits in $A$ followed by $B$, then $N_A + N_B - 1$ layers of FSWAP gates are required to do all the interactions by swapping the qubits in $A$ through $B$. Therefore all the fragment-bath interactions for one spin type can be done in $2N_\text{frag}-1$ layers and it can be shown that the fragment- and bath-only hopping terms can fit within these layers.

The more complicated case is when the shape of the fragment is also 2D. As before, there are nearest neighbour hopping terms in the fragment and each of the fragment sites on the edge interacts with all of the bath sites. However the bath-only hopping sites are split into four groups, rather than two. 

Let us take the fragment size to be $N_\text{frag} = N_x \times N_y$ and assume that $N_x \leq N_y$. The fragment sites will be ordered with the ``snake'' ordering~\cite{Cade2020, Jiang2018, Wecker2015_2} and will be followed by the bath sites placed in their four groups. All of the spin up sites will be followed by all spin down, but we restrict our analysis of the swap network to one spin type as before. We could use a different ordering, but this has the advantage of simplicity and enables us to use an efficient swap network for the fragment~\cite{Cade2020}.

The general structure of the swap network is as follows. The bath sites will be swapped through the fragment sites (where the bath sites come across a fragment edge site a combined FSWAP and hopping gate will be done, otherwise just an FSWAP) and simultaneously fragment edge sites will be moved along the snake towards the incoming bath sites. During this, the nearest neighbour hopping terms will be implemented using an efficient swap network designed for the Hubbard model~\cite{Cade2020}. The four sets of bath site hopping terms will be implemented using the full swap network~\cite{Kivlichan2018}. 

To determine the complexity of the circuit, we must consider the different components that make up the network separately. The fragment hopping terms (nearest neighbour horizontal and vertical) can be completed in $2N_x$ or $2N_x+1$ layers of two-qubit gates for $N_x$ even or odd respectively~\cite{Cade2020}. If the four groups of bath sites are of size $N_i \approx N_\text{frag}/4$ for $i=1,2,3,4$ then each of their individual swap networks can complete in $N_i$ layers~\cite{Kivlichan2018}. Finally, if all the $N_E$ fragment edge sites where $N_E = 2(N_x + N_y - 2)$ are placed next to the $N_\text{frag}$ bath sites then $N_\text{frag} + N_E - 1$ layers of two-qubit gates are required to carry out all the fragment-bath interactions. 

\begin{figure}[tb]
    \centering
    \subfloat[First two layers of the swap network for $N_x = 4$.]{\includegraphics[width=0.72\linewidth]{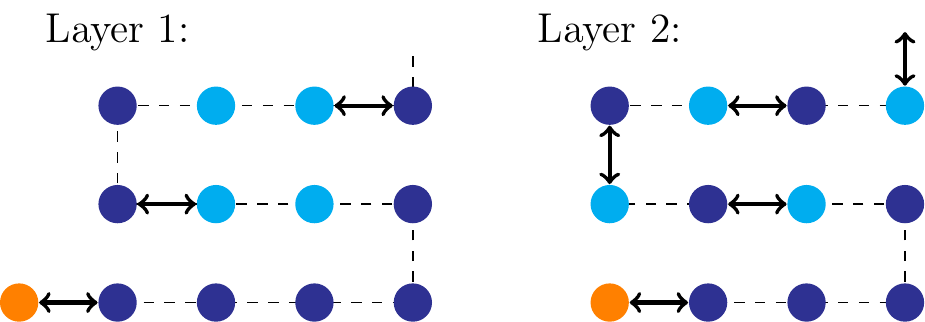} \label{fig:2Dswap_size4}}
    
    \subfloat[Selected layers of the swap network for $N_x = 6$.]{\includegraphics[width=\linewidth]{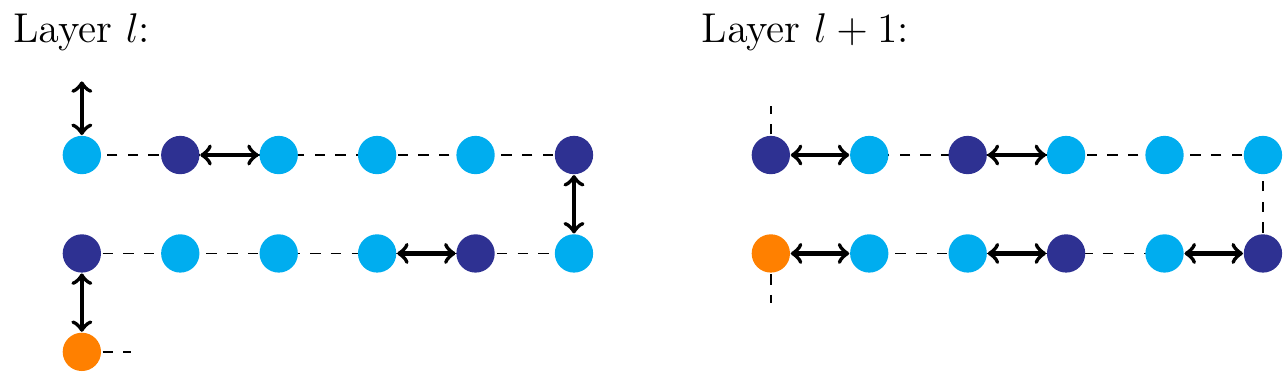} \label{fig:2Dswap_size6}}
    \caption{Sections of the swap network for a 2D fragment. The dark blue circles represent fragment edge sites, the light blue circles represent middle fragment sites and the orange circles are bath sites. The dotted line shows the Jordan-Wigner snake ordering; quantum gates can only happen along this line. When acting between a bath site and a fragment edge site, the arrows represent combined FSWAP and hopping gates; otherwise they represent FSWAP gates.}
    \label{fig:2Dswap_network}
\end{figure}

We now combine these three different swap networks together in as few layers as possible. Since the fragment-bath interactions will require the most layers, let us first consider how many will be required when the fragment starts in the snake ordering. For $N_x \leq 4$, it is possible to use FSWAPs to move the middle fragment sites away in time so that they never come into contact with the bath sites, as can be seen in Figure~\ref{fig:2Dswap_size4}. In this case all the fragment-bath interactions can be done in the minimum $N_\text{frag} + N_E - 1$ layers of two-qubit gates.

For $N_x > 4$, the pattern that the fragment sites settle into is: edge, middle, edge, $N_x - 3$ middle sites, repeated. When a bath site comes across this set of $N_x - 3$ middle sites, it is not possible to swap them away in time -- see Figure~\ref{fig:2Dswap_size6}. Every time this occurs an extra ${\ceil{(N_x-4)/2}}$ layers of FSWAPs is done to bring the bath site to the next fragment edge site. This happens $N_y - 4$ times during the entire swap network since the row of all fragment edge sites at the beginning and end of the fragment creates a buffer of edge sites. This leads to $(N_y - 4) \ceil{(N_x-4)/2}$ extra layers of two-qubit gates, as compared to the $N_x \leq 4$ case.

The next thing to check is whether the bath- and fragment-only interactions can fit within these layers, or if more will be required. Since $N_i \approx N_\text{frag}/4$, three of the bath swap networks can complete before they interact with the fragment sites, and the final one after its associated bath sites have passed through the fragment. 

This leaves the fragment hopping terms which require $2N_x$ layers to complete (taking $N_x$ to be even for this argument). In this time the bath sites can interact with all the sites in the first two rows of the fragment, which means that all the horizontal and vertical hopping terms for the third row of the fragment onwards can be implemented before they come into contact with bath sites. However, the vertical hopping terms between the first and second rows of the fragment will need to be carried out after they have swapped through the bath sites, requiring an extra $N_x - 2$ middle fragment sites to pass through all of the bath sites. This adds $N_x - 2$ layers to the swap network\footnote{Despite the fragment swap network for $N_x$ odd requiring an extra layer, this does not translate into an extra layer for the 2D embedded swap network. This is because the Jordan-Wigner snake ordering can be chosen such that the vertical hopping term that requires this extra layer is between the first two rows.}, making the final count of layers  
\begin{equation}
    L = N_\text{frag} + N_E + N_x - 3
\end{equation}
for $N_x \leq 4$, and 
\begin{equation}
    L + (N_y - 4)\ceil{\frac{N_x - 4}{2}}
\end{equation}
for $N_x > 4$.

At the next ansatz depth the swap network described here is reversed, allowing us to get the same circuit complexity at every depth.

\section{Generic measurement schemes for hopping terms}
\label{app:measurement}

In this appendix we describe two generic schemes for measuring hopping terms based on the measurement of non-crossing pairs described in Section~\ref{sec:measurement}.

\begin{figure*}[t]
    \centering
    \subfloat[]{\includegraphics[width=0.3\linewidth]{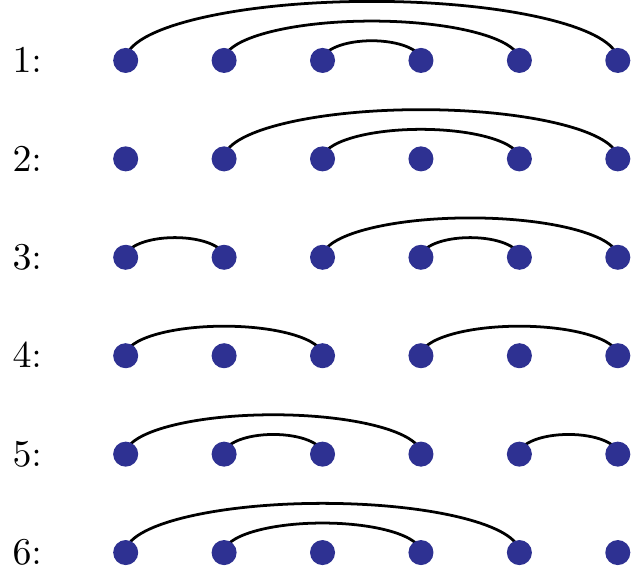} \label{fig:nc_measurements}} 
    \hspace{0.15\linewidth}
    \subfloat[]{\includegraphics[width=0.3\linewidth]{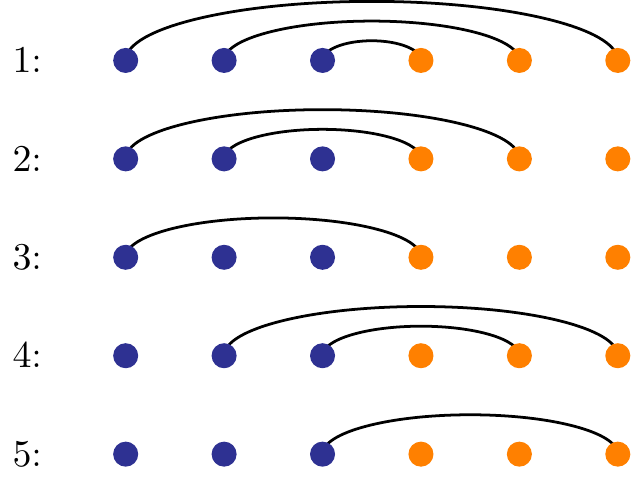} \label{fig:sets_measurements}} 
    \caption{(a) Measuring all hopping terms on $n$ qubits in $n$ rounds via rainbows of non-crossing pairs. Each row corresponds to a measurement round and each pair of qubits is included in exactly one round. (b) Measuring all hopping terms between qubits in set $A$ (blue) and qubits in set $B$ (orange) in $N_A+N_B-1$ circuit preparations.}
    \label{fig:measurement_schemes}
\end{figure*}

We first discuss how to measure hopping terms between every pair of qubits for an arbitrary number of qubits $n$ in $n$ preparations of the circuit. The idea is to produce a sequence of measurement rounds where each round contains at most two ``rainbows''. A rainbow between qubits $i$ and $j$ consists of the pairs
\begin{equation*}
    (i,j),(i+1,j-1),\dots,((i+j-1)/2,(i+j+1)/2)
\end{equation*}
when $j-i$ is odd, and the pairs
\begin{equation*}
    (i,j),(i+1,j-1),\dots,((i+j)/2-1,(i+j)/2+1)
\end{equation*}
when $j-i$ is even, corresponding to all non-crossing pairs between $i$ and $j$ centred at $(i+j)/2$. Then the $k^\text{th}$ round contains rainbows between qubits 1 and $k-1$, and between qubits $k$ and $n$ (where we do not include rainbows which have one end below qubit 1 or above qubit $n$). Each pair of qubits is then included in exactly one rainbow centred at their midpoint. This is illustrated for $n=6$ in Figure \ref{fig:nc_measurements}.

We remark that Hamamura and Imamichi~\cite{Hamamura2020} have developed a related procedure based on a heuristic algorithm that chooses pairs of qubits to measure in the Bell basis, in order to estimate the energy with respect to terms in an arbitrary Hamiltonian described by Pauli matrices. Many other techniques for reducing the number of measurement rounds required in VQE are known; see~\cite{Bharti2021} for a review.

We now turn to the situation where there are two sets $A$ and $B$ of $N_A \leq N_B$ qubits where every qubit in set $A$ shares a hopping term with every qubit in set $B$. Assuming the Jordan-Wigner ordering places $A$ before $B$, then all the hopping terms can be measured in $N_A + N_B - 1$ circuit preparations, saving one depth compared to the previous case. The measurement of the fragment-bath hopping terms is covered by this scenario.

At the first circuit preparation we measure all of the qubits in $A$ with the furthest $N_A$ qubits in $B$. At subsequent preparations we make our way down through the qubits in $B$ until the first qubit in $A$ has been measured with the first qubit in $B$. This requires $N_B$ circuit preparations. For the remaining $N_A-1$ preparations, we switch and measure the last qubits in $B$ with the furthest possible qubits in $A$ (that have not already been measured). This is shown in Figure~\ref{fig:sets_measurements} for $N_A = N_B = 3$.

\bibliographystyle{mybibstyle}
\bibliography{bibliography}

\end{document}